\documentclass[preprint2]{aastex}
 \usepackage[fleqn]{amsmath}
 \usepackage{graphicx}
\usepackage{dcolumn}
\usepackage{bm}
\usepackage{multirow}

\shorttitle{Construction of explicit symplectic integrators}
\shortauthors{Wang et al.}

\begin{document}


\title{Construction of explicit symplectic integrators in general relativity. II. Reissner-Nordstr\"{o}m black holes}


\author{Ying Wang$^{1,2}$, Wei Sun$^{1}$, Fuyao Liu$^{1}$, Xin Wu$^{1,2,3,\dag}$}
\affil{1. School of Mathematics, Physics and Statistics, Shanghai
University of Engineering Science, Shanghai 201620, China
\\ 2. Center of Application and Research of Computational Physics,
Shanghai University of Engineering Science, Shanghai 201620, China
\\  3. Guangxi Key Laboratory for Relativistic Astrophysics, Guangxi
University, Nanning 530004, China} \email{Emails:
wangying424524@163.com (Y. W.), sunweiay@163.com (W. S.),
liufuyao2017@163.com (F. L.); $\dag$ Corresponding Author:
wuxin$\_$1134@sina.com (X. W.)}


\begin{abstract}

In a previous paper, second- and fourth-order explicit symplectic
integrators were designed for a Hamiltonian of the Schwarzschild
black hole. Following this work, we continue to trace the
possibility of the construction of explicit symplectic integrators
for a Hamiltonian of charged particles moving around a
Reissner-Nordstr\"{o}m black hole with an external magnetic field.
Such explicit symplectic methods are still available when the
Hamiltonian is separated into five independently integrable parts
with analytical solutions as explicit functions of proper time.
Numerical tests show that the proposed algorithms share the
desirable properties in their  long-term  stability, precision and
efficiency for appropriate choices of step sizes. For the
applicability of one of the new algorithms, the effects of the
black hole's charge, the Coulomb part of the electromagnetic
potential and the magnetic parameter on the dynamical behavior are
surveyed.  Under some circumstances, the extent of chaos gets
strong with an increase of the magnetic parameter from a global
phase-space structure. No the variation of the black hole's charge
but the variation of the Coulomb part is considerably sensitive to
affect the regular and chaotic dynamics of particles' orbits. A
positive Coulomb part is easier to induce chaos than a negative
one.

\end{abstract}


\emph{Unified Astronomy Thesaurus concepts}: Black hole physics
(159); Computational methods (1965); Computational astronomy
(293); Celestial mechanics (211)




\section{Introduction}
\label{sec:intro}

Based on Einstein's theory of general relativity, black holes are
a group of solutions of the Einstein's field equations. Usually,
black holes have singularities covered by event horizon surfaces.
The first black hole's solution for the description of a static
and spherically symmetric gravitational field around a point-like
mass was given by Schwarzschild (1916). When the mass source in
the origin is charged, the Reissner-Nordstr\"{o}m (RN) metric
(Reissner 1916) became available. A rigorous solution representing
the gravitational field around a rotating central mass is the Kerr
metric (Kerr 1963). On the other hand, evidence from observations
demonstrated the existence of supermassive black holes with masses
from millions to tens of billions of solar masses in centers of
nearly all galaxies. In terms of images of M87, a central Kerr
black hole is estimated to have mass $M=(6.5\pm 0.7)\times
10^{9}M_{\odot}$ ($M_{\odot}$ being the Sun's mass), which is
consistent with the result predicted by the general theory of
relativity (EHT Collaboration et al. 2019a, b, c). Successful
gravitational-wave measurements (Abbott et al. 2016) also provided
powerful evidence for the presence of black holes.

Although the relativistic spacetimes like the RN or Kerr metric
are highly nonlinear, they are integrable and have analytical
solutions because of the presence of enough constants of motion.
The solutions have only formal expressions in terms of
quadratures, but cannot be expressed as elementary functions or
explicit functions of time. To know detailed information on the
solutions how to evolve with time, one had better employ a
numerical integration technique to solve the integrable problems.
If the central bodies are suffered from perturbations, such as
external magnetic fields, the spacetimes become non-integrable in
most cases. Under some circumstances, chaos occurs (Takahashi $\&$
Koyama 2009; Kop\'{a}\v{c}ek et al. 2010; Kop\'{a}\v{c}ek $\&$
Karas 2014; Stuchl\'{i}k $\&$ Kolo\v{s} 2016; P\'{a}nis et al.
2019; Li $\&$ Wu 2019; Stuchl\'{i}k et al. 2020; Yi $\&$ Wu 2020).
This chaoticity indicates that a dynamical system is exponentially
sensitive dependence on initial conditions (Lichtenberg $\&$
Lieberman 1983). In this case, the numerical technique is more
indispensable to study the non-integrable systems.

Reliable results from numerical integrators with a good behavior
are required, especially in the case of long-term integration of
chaotic orbits. The most appropriate solvers are geometric or
structure preserving algorithms (Hairer et al. 1999; Seyrich $\&$
Lukes-Gerakopoulos 2012; Bacchini et al. 2018a, 2018b; Hu et al.
2019), such as symplectic methods for Hamiltonian systems (Ruth
1983; Wisdom $\&$ Holman 1991). They have several advantages over
standard explicit integrators, such as the family of explicit
Runge-Kutta solvers. The integrals of motion (e.g., energy
integral) along the trajectory are nearly conserved for the
structure preserving integrators, but their errors increase
linearly with time for the standard integration schemes. In
addition, the overall phase error only grows linearly with time
for the former algorithms, whereas it is normally proportional to
the square of the length of the integration interval for the
latter schemes (Deng et al. 2020). The above-mentioned curved
spacetimes can be expressed in terms of Hamiltonian systems, and
thus symplectic integrators are naturally chosen. Standard
explicit symplectic integrators, such as a second-order Verlet
integrator (Swope et al. 1982), become useless. However,
completely implicit symplectic methods (Kop\'{a}\v{c}ek et al.
2010; Seyrich $\&$  Lukes-Gerakopoulos 2012; Tsang et al. 2015),
such as the implicit midpoint scheme (Feng 1986; Brown 2006), or
implicit and explicit combined symplectic methods (Liao 1997;
Preto $\&$ Saha 2009; Lubich et al. 2010; Zhong et al. 2010; Mei
et al. 2013a, 2013b) are used. This is because the Hamiltonians
have no separable forms of variables or can be split into two
integrable parts without analytical solutions as explicit
functions of time. Unfortunately, such implicit integrators  are
more computationally demanding  at the expense of computational
time than the same order standard explicit methods. Extended
phase-space methods (Pihajoki 2015; Liu et al. 2016; Luo et al.
2017; Li $\&$ Wu 2017) are explicit and have good long-term stable
behavior in energy errors, but are not symplectic. In a previous
work (Wang et al. 2021), we overcame the difficulty in the
construction of explicit symplectic integrators for the
Schwarzschild metric. In our construction, the Hamiltonian for the
Schwarzschild spacetime can be separated into four integrable
parts with analytical solutions as explicit functions of proper
time. Then, these explicit solvable operators symmetrically
composed second- and fourth-order explicit symplectic integrators.

Following the previous work (Wang et al. 2021), we design explicit
symplectic integrators for the RN black hole immersed into an
external magnetic field. This is one of the main aims in the
present paper. Another aim is to know how  an increase of the
black hole's charge, the Coulomb paramter of the electromagnetic
potential or the magnetic parameter exerts an influence on the
dynamical transition of orbits of charged particles around the RN
black hole. For the sake of these purposes, we introduce a
dynamical model of charged particles moving around the RN black
hole surrounded with an external magnetic field in Section 2.
Second- and fourth-order explicit symplectic integrators are
designed for the magnetized RN spacetime in Section 3. In Section
4, we evaluate the numerical performance of the proposed
algorithms, and apply a new integrator to address the question of
how the related parameters affect the orbital dynamics of order
and chaos. Finally, the main results are concluded in Section 5.

\section{Reissner-Nordstr\"{o}m black holes}

The Schwarzschild black hole with charge $Q$  is the RN black
hole. In dimensionless spherical-like coordinates $(t, r, \theta,
\phi)$, the RN spacetime (Reissner 1916) takes the following
metric
\begin{eqnarray}
-\tau^2 &=& ds^{2} = g_{\alpha\beta}dx^{\alpha}dx^{\beta} \nonumber \\
&=& -(1-\frac{2}{r}+\frac{Q^2}{r^2})dt^{2}
+(1-\frac{2}{r}+\frac{Q^2}{r^2})^{-1} dr^{2}
\nonumber \\
& &+r^{2}d \theta^{2} +r^{2}\sin^{2} \theta d \phi^2.
\end{eqnarray}
The speed of light  $c$ and the constant of gravity $G$ use
geometrized units, $c=G=1$. $M$ is the mass of black hole,  and
also takes one unit, $M=1$. Proper time $\tau$, coordinate time
$t$, radial separation $r$ and charge $Q$ are dimensionless. In
practice, the dimensionless operations are obtained via scale
transformations: $\tau\rightarrow M\tau$, $t\rightarrow Mt$,
$r\rightarrow Mr$ and $Q\rightarrow MQ$. When $|Q|<1$, this
spacetime corresponds to black holes with two event horizons
$r_{\pm}=1\pm\sqrt{1-Q^2}$. The spacetime is still a black hole
with an event horizon $r=1$ for $Q=\pm 1$. It has no event horizon
but has naked singularities if $|Q|>1$. Hereafter, the case of
black holes with $|Q|\leq1$ is considered.

The motion of a test particle around the black hole is described
by the Lagrangian system
\begin{equation}
\ell = \frac{1}{2} (\frac{ds}{d\tau})^2
=\frac{1}{2}g_{\mu\nu}\dot{x}^{\mu}\dot{x}^{\nu},
\end{equation}
where $\dot{x}^{\mu}=\mathbf{U}$ is a four-velocity satisfying the
relation
\begin{equation}
\mathbf{U}\cdot\mathbf{U}=U^{\alpha}U_{\alpha}=g_{\mu\nu}\dot{x}^{\mu}\dot{x}^{\nu}=-1.
\end{equation}
A covariant generalized momentum $\mathbf{p}$ is defined as
\begin{equation}
p_{\mu} = \frac{\partial \ell}{\partial
\dot{x}^{\mu}}=g_{\mu\nu}\dot{x}^{\nu}.
\end{equation}
Because $t$ and $\phi$ do not explicitly appear in the Lagrangian,
there are two constant momentum components
\begin{eqnarray}
p_{t} &=& -(1-\frac{2}{r}+\frac{Q^2}{r^2})\dot{t}=-\mathcal{E},\\
p_{\phi} &=& r^{2}\sin^{2}\theta\dot{\phi}=\mathfrak{L}.
\end{eqnarray}
$\mathcal{E}$ and $\mathfrak{L}$ denote the particle's energy and
angular momentum, respectively.

The Lagrangian corresponds to the Hamiltonian
\begin{eqnarray}
H &=& \mathbf{U}\cdot\mathbf{p}-\ell
=\frac{1}{2}g^{\mu\nu}p_{\mu}p_{\nu} \nonumber \\
&=& -\frac{\mathcal{E}^{2}}{2}(1-\frac{2}{r}+\frac{Q^2}{r^2})^{-1}
+\frac{p^{2}_{r}}{2}(1-\frac{2}{r}+\frac{Q^2}{r^2}) \nonumber \\
&& +\frac{1}{2}\frac{p^{2}_{\theta}}{r^2}
+\frac{1}{2}\frac{\mathfrak{L}^{2}}{r^2\sin^2\theta}.
\end{eqnarray}
Because of the four-velocity  relation (3), this Hamiltonian is
always identical to -1/2,
\begin{equation}
H=-\frac{1}{2}.
\end{equation}
By separating the variables in the Hamilton-Jacobi equation, one
can find a second integral excluding the two integrals
$\mathcal{E}$ and $\mathfrak{L}$ in the Hamiltonian system (Carter
1968). Thus, this system is integrable and has formally analytical
solutions.

Now, suppose the black hole  surrounded by an external magnetic
field whose four-vector potential has two nonzero covariant
components
\begin{equation}
A_t=-\frac{Q}{r}, ~~~
A_{\phi}=\frac{B}{2}g_{\phi\phi}=\frac{B}{2}r^{2}\sin^{2} \theta,
\end{equation}
where $A_t$ represents the Coulomb part of the electromagnetic
potential (Kopaccek $\&$ Karas 2014), and $B$ is the strength of
the magnetic field parallel to the $z$ axis (Felice $\&$ Sorge
2003). The motion of a particle with charge $q$ under the
interactions of the black hole's gravity and electromagnetic force
is described by the Hamiltonian
\begin{eqnarray}
K &=& \frac{1}{2}g^{\mu\nu}(p_{\mu}-qA_{\mu})(p_{\nu} -qA_{\nu})
\nonumber \\
&=& -\frac{1}{2}(1-\frac{2}{r}+\frac{Q^2}{r^2})^{-1}
(E-\frac{Q^{*}}{r})^{2} \nonumber \\
&& +\frac{1}{2}(1-\frac{2}{r} +\frac{Q^2}{r^2})p^{2}_{r}
+\frac{1}{2}\frac{p^{2}_{\theta}}{r^2} \nonumber \\
&& +\frac{1}{2r^2\sin^2\theta}(L-\frac{1}{2}\beta r^{2}\sin^{2}
\theta)^{2},
\end{eqnarray}
where $Q^{*}=qQ$ is a Coulomb parameter of the electromagnetic
potential, and $\beta=qB$. To make the system (10) be
dimensionless, we take $K\rightarrow m^2K$, $E\rightarrow mE$,
$p_r\rightarrow mp_r$, $p_{\theta}\rightarrow mMp_{\theta}$,
$L\rightarrow mML$, $q\rightarrow mq$ and $B\rightarrow B/M$,
where $m$ is the particle's mass. The expressions of energy $E$
and angular momentum $L$ of the charged particle become
\begin{eqnarray}
E &=& (1-\frac{2}{r}+\frac{Q^2}{r^2})\dot{t}+\frac{Q^{*}}{r},
\\
 L &=& r^{2}\sin^{2}\theta\dot{\phi}+\frac{1}{2}\beta
r^{2}\sin^{2} \theta.
\end{eqnarray}
Similar to $H$, $K$ always satisfies the constraint
\begin{equation}
K=-\frac{1}{2}.
\end{equation}
However, $K$ unlike $H$ has no second integral. Thus, it is
non-integrable and has no formally analytical solutions. In this
case, a numerical integration method is a convenient tool to work
out such a non-integrable system.

\section{Construction of explicit symplectic integrators}

In view of a symplectic integrator with good geometric and
physical properties, it is naturally a prior choice of numerical
integrator for the description of long-term qualitative evolution
of the Hamiltonian system (7). An explicit symplectic  method
becomes useless without doubt if this Hamiltonian is separated
into two analytically solvable parts. This is because not all
analytical solutions of the two splitting parts are explicit
functions of proper time $\tau$. An explicit symplectic algorithm
fails to be built if the Hamiltonian is split into four
analytically integrable parts, as the Hamiltonian of Schwarzschild
black hole is in our previous paper (Wang et al. 2021). Thus, it
may be necessary that more analytically integrable splitting parts
should be given to the Hamiltonian for the construction of
explicit symplectic schemes.

Let the Hamiltonian of RN black hole be separated into five
separable parts
\begin{equation}
H=H_1+H_2+H_3+H_4+H_5,
\end{equation}
where the five sub-Hamiltonians are written as follows:
\begin{eqnarray}
H_1 &=&\frac{1}{2}\frac{\mathfrak{L}^{2}}{r^2\sin^2\theta}
-\frac{\mathcal{E}^{2}}{2}(1-\frac{2}{r}+\frac{Q^2}{r^2})^{-1}, \\
H_{2} &=& \frac{1}{2}p^{2}_{r},\\
H_{3} &=& -\frac{1}{r}p^{2}_{r},\\
H_{4} &=& \frac{p^{2}_{\theta}}{2r^2}, \\
H_{5} &=& \frac{1}{2}\frac{Q^2}{r^2}p^{2}_{r}.
\end{eqnarray}
$H_2$, $H_3$ and $H_4$ are the same as those in the Hamiltonian
splitting of Schwarzschild black hole  in the previous work (Wang
et al. 2021).

$H_1$ has its canonical equations $\dot{r}=\dot{\theta}=0$ and
\begin{eqnarray}
\frac{dp_{r}}{d\tau} &=& -\frac{\partial H_1}{\partial r} =
\frac{\mathfrak{L}^{2}}
{r^3\sin^2\theta}-\frac{\mathcal{E}^{2}}{r^{2}}(1-\frac{Q^{2}}{r})
\nonumber
\\ && \cdot  (1-\frac{2}{r}+\frac{Q^2}{r^2})^{-2}=\Re(r,\theta),\\
\frac{dp_{\theta}}{d\tau} &=& -\frac{\partial H_1}{\partial
\theta} = \frac{\mathfrak{L}^2\cos\theta}
{r^{2}\sin^{3}\theta}=\Theta(r,\theta).
\end{eqnarray}
If $\mathcal{A}$ is taken as a differential operator
\begin{eqnarray}
\mathcal{A} = \Re\frac{\partial }{\partial
p_r}+\Theta\frac{\partial }{\partial p_{\theta}},
\end{eqnarray}
then  $\mathcal{A}p_r=\dot{p}_r=\Re$ and
$\mathcal{A}p_{\theta}=\dot{p}_{\theta}=\Theta$. Because $r$ and
$\theta$ are constants, $p_r$ and $p_{\theta}$ are easily solved.
From proper time $\tau_0$ over a proper time step $h$ to proper
time $\tau=\tau_0+h$, the solutions are expressed as
\begin{eqnarray}
p_{r} &=& p_{r0} +h\Re(r_0,\theta_0),\\
p_{\theta} &=& p_{\theta0}+h\Theta(r_0,\theta_0),
\end{eqnarray}
where $\textbf{z}(0)=(r_0, \theta_0, p_{r0},  p_{\theta0})$ are
the solutions at the beginning of the step of length $h$. We use
an exponential operator $e^{h\mathcal{A}}$ to represent the
analytical solutions (23) and (24), i.e., $(p_{r}, p_{\theta}) =
e^{h\mathcal{A}}\textbf{z}(0)$.

Set $\mathcal{B}$, $\mathcal{C}$, $\mathcal{D}$ and $\mathcal{F}$
as differential operators of $H_2$, $H_3$, $H_4$ and $H_5$,
respectively. They are of the following expressions
\begin{eqnarray}
\mathcal{B} &=& p_{r}\frac{\partial}{\partial r}, \\
\mathcal{C} &=& -\frac{2}{r}p_r\frac{\partial}{\partial r}
-\frac{p^{2}_r}{r^2}\frac{\partial}{\partial p_r}, \\
\mathcal{D} &=& \frac{p_{\theta}}{r^{2}}\frac{\partial}{\partial
\theta}+\frac{p^2_{\theta}}{r^{3}}\frac{\partial}{\partial p_r}, \\
\mathcal{F} &=&  \frac{Q^2}{r^2}p_r\frac{\partial}{\partial
r}+Q^2\frac{p^{2}_r}{r^3}\frac{\partial}{\partial p_r}.
\end{eqnarray}
The four sub-Hamiltonians have their analytical solutions
\begin{eqnarray}
e^{h\mathcal{B}}: ~r &=& r_0+h p_{r0}; \\
e^{h\mathcal{C}}: ~ r &=& [(r^{2}_{0}-3h
p_{r0})^{2}/r_0]^{1/3}, \nonumber \\
 p_r &=& p_{r0}[(r^{2}_{0}-3h
p_{r0})^{2}/r^2_0]^{1/3}; \\
e^{h\mathcal{D}}: ~ \theta &=& \theta_0+h p_{\theta0}/r^{2}_{0}, \nonumber \\
 p_r &=& p_{r0}+h p^2_{\theta0}/r^{3}_{0}; \\
e^{h\mathcal{F}}: ~ r &=& \sqrt{r^{2}_{0}+2hQ^2p_{r0}/r_0}, \nonumber \\
 p_r &=& \frac{p_{r0}}{r_0}\sqrt{r^{2}_{0}+2hQ^2p_{r0}/r_0}.
\end{eqnarray}
It is clear that all the analytical solutions in Equations (23),
(24) and (29)-(32) are explicit functions of proper time $\tau$ or
step size $h$. Although the  compositions $H_2+H_3$, $H_2+H_3+H_4$
and $H_2+H_3+H_4+H_5$ can be solved analytically, their solutions
are not expressed in terms of  explicit functions of $\tau$. The
present splitting form (14) of the Hamiltonian $H$ is one possible
choice to satisfy the need.

The solutions of the Hamiltonian (7) over the time step $h$ can be
obtained approximately by a second order explicit symplectic
integrator, namely, symmetric products of these exponential
operators
\begin{eqnarray}
S^{H}_2(h) &=& e^{\frac{h}{2}\mathcal{F}}
e^{\frac{h}{2}\mathcal{D}} e^{\frac{h}{2}\mathcal{C}}
e^{\frac{h}{2}\mathcal{B}} e^{h\mathcal{A}} \nonumber \\
&& \otimes e^{\frac{h}{2}\mathcal{B}} e^{\frac{h}{2}\mathcal{C}}
e^{\frac{h}{2}\mathcal{D}} e^{\frac{h}{2}\mathcal{F}}.
\end{eqnarray}
It can compose a fourth-order symplectic scheme of Yoshida (1990)
\begin{equation}
S^{H}_4(h)=S^{H}_2(\gamma h)\circ S^{H}_2(\delta h)\circ
S^{H}_2(\gamma h),
\end{equation}
where $\delta=1-2\gamma$ and $\gamma=1/(2-\sqrt[3]{2})$.

The two explicit symplectic algorithms for $H$ are also suitable
for $K$. The only one difference is $H_1$ replaced with
\begin{eqnarray}
K_1 &=& \frac{1}{2r^2\sin^2\theta}(L-\frac{1}{2}\beta
r^{2}\sin^{2}
\theta)^{2} \nonumber \\
&& -\frac{1}{2}(1-\frac{2}{r}+\frac{Q^2}{r^2})^{-1}
(E-\frac{Q^{*}}{r})^{2}.
\end{eqnarray}
Then, we obtain two explicit symplectic methods $S^{K}_{2}$ and
$S^{K}_{4}$ for the Hamiltonian $K$.

\begin{table*}[htbp]
\centering \caption{Performance of algorithms S2 and S4 with
different time steps $h$. In the brackets, e.g., ($10^{-8}$, U,
$11'40''$), $10^{-8}$ denotes the order of Hamiltonian error, U
(or B) indicates the unboundedness (or boundedness) of Hamiltonian
error, and $11'40''$ corresponds to CPU time (minute $'$, second
$''$). The integration time reaches $\tau=10^{8}$ for each step
size.} \label{Tab1}
\begin{tabular}{lccccccccc}
\hline h   & 0.1   & 1  &  4  & 10\\
\hline S2  & ($10^{-8}$, U, $11'40''$)     & ($10^{-6}$, B, $1'12''$)  & ($10^{-5}$, B, $20''$)   & ($10^{-4}$, B, $7''$) \\
\hline S4  & no tested                     & ($10^{-9}$, U, $3'34''$)  & ($10^{-8}$, B, $56''$)   & ($10^{-6}$, B, $23''$) \\
\hline
\end{tabular}
\end{table*}

\section{Numerical simulations}

At first, let us check the numerical performance of the proposed
explicit symplectic integration algorithms for solving the system
(10). Then, one of the new methods is selected to explore the
orbital dynamics of charged massless particles in the system.

\subsection{Evaluations of the new algorithms}

In the previous work (Wang et al. 2021), the established explicit
symplectic integrators for the Schwarzschild black hole surrounded
by an external magnetic field were compared with a conventional
fourth-order Runge-Kutta integrator, second- and fourth-order
explicit and implicit mixed symplectic algorithms (Mei et al.
2013b) and second- and fourth-order extended phase-space explicit
symplectic-like methods (Luo et al. 2017). It was shown that the
Runge-Kutta method has a secular drift in Hamiltonian errors and
performs the poorest performance. The other algorithms at same
order can exhibit good long-term stable error behavior for
appropriate time steps and have no explicit differences among
their Hamiltonian errors. Therefore, only the newly proposed
explicit symplectic integrators in the present paper are
considered to work out the Hamiltonian $K$.

Taking proper time step $h=1$, we consider the parameters to be
$E=0.995$, $L=4.6$, $\beta=6.4\times10^{-4}$, $Q=0.1$ and
$Q^{*}=10^{-4}$.  The new second-order explicit symplectic
integrator S2 (or the new fourth-order method S4) is used to
integrate an orbit with initial conditions $r=25$, $p_r=0$ and
$\theta=\pi/2$. The starting value of  $p_{\theta}>0$ is
determined by Equation (13). In Figure 1(a), Hamiltonian errors
$\Delta K=-1/2-K$ in Equation (13) can remain bounded in an order
of $\mathcal{O}(10^{-6})$ for S2 when the number of integration
steps is $10^{8}$. S4 gives a higher accuracy with an order of
$\mathcal{O}(10^{-9})$, but its errors grow linearly with time
due to roundoff errors. Here are some analysis to these results.
The test orbit has an approximate average period $T\approx 5000$.
Truncation Hamiltonian error is $(h/T)^2\sim10^{-8}$ for S2 and
$(h/T)^4\sim10^{-16}$ for S4. In fact, the error outputted at the
end of the first step is $1.98\times 10^{-15}$ for S4. In
addition, the machine yields a roundoff error in per computation,
e.g. $\epsilon=10^{-16}$ in a double-precision level. The roundoff
errors grow in a rough estimation $n\epsilon$, where $n$ is a
number of computations. The roundoff errors are more important
than the truncation errors when $n$ is large enough. In an
integration time $t=200$, the Hamiltonian errors for S4 fast grow
to $9\times10^{-12}$. When $t$ spans this time and is less than
$10^{7}$, the errors much slowly grow and basically remain stable
at an order of $\mathcal{O}(10^{-9})$. With the integration
continuing, the boundness of the Hamiltonian errors is destroyed
by the roundoff errors. If the step size gets larger, e.g. $h$=10,
the Hamiltonian errors are stabilized at an order of
$\mathcal{O}(10^{-6})$. For $h$=4, the Hamiltonian errors remain
bounded in an order of $\mathcal{O}(10^{-8})$. These results
roughly indicate that a symplectic integrator can stabilize the
Hamiltonian errors at the values larger than
$\mathcal{O}(10^{-8})$ for $10^{8}$ integration steps, which yield
roundoff errors in the order of $\mathcal{O}(10^{-8})$. When the
time step is $h=1$ and the number of integration steps is
$10^{8}$, the main error source for S2 is the truncation errors
and therefore the Hamiltonian errors can remain stable at the
order of $\mathcal{O}(10^{-6})$. However, the roundoff errors for
S2 with time step $h=0.1$ reach an order of $\mathcal{O}(10^{-7})$
after $10^{9}$ integration steps. This forces the Hamiltonian
errors with an order of $\mathcal{O}(10^{-8})$ to grow linearly.
To clearly show the dependence of the magnitude and boundness of
Hamiltonian errors and computational efficiency for algorithms S2
and S4 on the time step $h$, we give Table 1. Obviously, S2 with
$h=1$ or S4 with $h=4$ is an optimal choice in the present cases.

The test orbit in Figure 1(a) is Orbit 1 colored red in Figure
1(b). Because this orbit is a single Kolmogorov-Arnold-Moser (KAM)
torus on the Poincar\'{e} section, it is a regular quasi-periodic
orbit. Black Orbit 2 with the initial value $r=90$, consisting of
11 small islands, is also a regular many-islands KAM torus. The
occurrence of resonance and chaos will become easy for such an
orbit with many islands. However, blue Orbit 3 with the initial
value $r=50$ has many discrete points distributed in a small area
on the Poincar\'{e} section. This kind of phase space structure
indicates the chaoticity of Orbit 3. The purple orbit with the
initial value $r=110$ is also chaotic. In spite of the onset of
chaos, the possibility of ``islands of regularity" like Orbits 1
and 2 is still existent. Based on KAM theorem, the minima of the
effective potential in the equatorial plane correspond to stable
circular orbits, which are related to regular harmonic oscillatory
motions for the description of Keplerian accretion disks of
stellar mass black holes. Moreover, all trajectories that are
bounded in the vicinity of the equatorial plane are also regular
(Kolo\v{s} et al. 2015). These regular motions can successfully
explain the quasi-periodic oscillations of X-ray flux from several
microquasars (Kolo\v{s} et al. 2017; Tursunov $\&$ Kolo\v{s}
2018). There is another island of regularity related to the motion
along the magnetic field lines (Tursunov et al. 2020a). In some
cases, a hot spot can exhibit quasi-circular motion along a single
orbit (Tursunov et al. 2020b). However,  the quasicircular  motion
may become chaotic because  the axial symmetry of the system is
broken so that the hot spot's angular momentum is not conserved.
The absence of the angular momentum is caused by the inclination
angle of the hot-spot orbit from the equatorial plane or of the
magnetic field lines with respect to the black hole's spin axis.

If regular single-island Orbit 1 is replaced with regular
many-islands Orbit 2 or chaotic Orbit 3, the numerical performance
of the two algorithms S2 and S4 has no explicit differences. In
other words, no dynamical behavior of orbits but a step size
mainly affects the quality of the proposed algorithms. In the
later discussions, we employ S4 with the appropriate time step
$h=4$ to investigate the related dynamical features of the
Hamiltonian $K$ when charge parameters $Q$ and $Q^{*}$, and
magnetic parameter $\beta$ are varied.

\subsection{Applications}

To show the dependence of the orbital dynamics of order and chaos
on the black hole's charge $Q$ or the Coulomb parameter $Q^{*}$,
we fix the parameters $E$, $L$ and $\beta$ in Figure 1. Of course,
different values of $Q$ and $Q^{*}$ are adopted.

In fact, the phase-space structures for the case of $Q=Q^{*}=0$
are similar to those for the case of $Q=0.1$ and $Q^{*}=10^{-4}$
in Figure 1(b). To clearly show how the orbital dynamics of order
and chaos depends on the charge $Q$, we consider the choice of
$Q\gg Q^{*}$. Fixing $Q^{*}=10^{-4}$, we give $Q$ different larger
values. The result for $Q=0.1$ is also suitable for the case of
$Q=0.3$. However, the phase-space structures for $Q=0.6$ in Figure
2(a) are somewhat different from those for $Q=0.1$. Orbit 2 is
ordered in Figure 1(b), but becomes chaotic in Figure 2(a). Orbit
3 that is chaotic in Figure 1(b) is a regular orbit with many
loops in Figure 2(a). Compared with those for the case of $Q=0.6$
in Figure 2(a), the orbits exist some differences for the case of
$Q=0.8$ in Figure 2(b). The blue islands in Figure 2(a) become an
ordered single torus in Figure 2(b). The green ordered single
torus in Figure 2(a) is weakly chaotic in Figure 2(b). When $Q=1$,
the extent of chaos in Figure 2(c) is not typically strengthened.
A result seems to be concluded from Figures 1(b) and 2(a)-2(c). An
increase of the black hole's charge $Q$ may exert some influence
on the phase-space structures, but does not bring an apparent
dynamical transition from order to chaos. Namely, it is not
considerably sensitive to alter the dynamical orbital properties.
It does not typically enhance the extent of chaos, either.

What about the dynamical transition with an increase of the
Coulomb parameter $Q^{*}$ for a given smaller value (e.g.,
$Q=10^{-4}$)? Figures 2(d) and 2(e) describe that the chaotic
behavior existing in the case of $Q=Q^{*}=0$ gradually dies out
when $Q^{*}$ increases, such as $Q^{*}=$0.1 and 0.3. As $Q$ runs
from a smaller value $Q=10^{-4}$ to a larger value $Q=0.3$, the
phase-space structures have no dramatic differences between
Figures 2(f) and 2(e). In fact, the phase-space structures for the
case of $Q=Q^{*}=0.1$ are basically similar to those for the case
of $Q=10^{-4}$ and $Q^{*}=0.1$ in Figure 2(d). The orbits for the
case of $Q=0.1$ and $Q^{*}=0.3$ are also the same as those for the
case of $Q=10^{-4}$ and $Q^{*}=0.3$ in Figure 2(e). This result
shows again that no $Q$ but $Q^{*}$ mainly affects the regular and
angular dynamics of orbits. In particular, a positive value of
$Q^{*}$ weakens the strength of chaos. On the other hand, a
negative value of $Q^{*}$ can easily induce the occurrence of
chaos, and the extent of chaos is drastically strengthened when
the magnitude of negative Coulomb parameter $Q^{*}$ increases, as
shown in Figures 2(g)-2(i). Notice that the orbits between the
case of $Q=0.1$ and $Q^{*}=-0.1$ and the case of $Q=10^{-4}$ and
$Q^{*}=-0.1$ in Figure 2(g) are almost the same. So are the orbits
between the case of $Q=0.1$ and $Q^{*}=-0.3$ and the case of
$Q=10^{-4}$ and $Q^{*}=-0.3$ in Figure 2(i). When Coulomb
parameter $Q^{*}$ is negative, the Coulomb part of the
electromagnetic potential $A_t$ in Equation (9) is positive.

What will happen if magnetic parameter $\beta$ increases but
parameters $E$, $L$, $Q$ and $Q^{*}$ are fixed? Red Orbit 1 with
parameters $Q=0.1$ and $Q^{*}=10^{-4}$ in Figure 1(b) is tested.
The orbit is twisted for $\beta=9.7\times 10^{-4}$ in Figure 3(a),
becomes a three-islands orbit for $\beta=9.9\times 10^{-4}$ in
Figure 3(b), and is finally evolved to a strong chaotic orbit for
$\beta=1.1\times 10^{-3}$ in Figure 3(c). Given $Q=10^{-4}$ and
$Q^{*}=0.1$, the orbit becomes many-islands, weakly chaotic and
strong chaotic orbits as $\beta$ increases from $9.7\times
10^{-4}$ to $1.1\times 10^{-3}$ in Figures 3(d)-3(f). For
$Q=Q^{*}=0.1$, the orbit is  evolved to a twisted single torus, a
many-islands orbit and a strong chaotic orbit with an increase of
$\beta$  in Figures 3(g)-3(i).  Particular for $Q^{*}=$-0.1, -0.3
and $\beta=9.7\times 10^{-4}$, $9.9\times 10^{-4}$, $1.1\times
10^{-3}$, strong chaos (not plotted) always occurs. All the
results prove that an increase of magnetic parameter $\beta$ gives
rise to enhancing the chaotic effect.

Why do the two charge parameters $Q$ and $Q^{*}$ have completely
different effects on the dynamical behavior of orbits? Why does an
increase of negative parameter $Q^{*}$ or magnetic parameter
$\beta$ lead to strengthening the extent of chaos? To answer these
questions, we expand the term $(~)^{-1}$ in Equation (35) and
rewrite Equation (35) as follows:
\begin{eqnarray}
K_1 &\approx& -\frac{1}{2}(\beta L+E^2)+\frac{\beta^2}{8}
r^{2}\sin^{2} \theta -\frac{E^2}{r} \nonumber \\
&& +\frac{EQ^{*}}{r}+
\frac{L^2}{2r^2\sin^2\theta}+\frac{Q^2E^2}{2r^2} \nonumber \\
&& +\frac{Q^{*}}{2r^2}(4E-Q^{*})+\cdots.
\end{eqnarray}
The second term $V_1=\beta^2r^{2}\sin^{2} \theta/8$ in Equation
(36) is a magnetic field force acting as a gravitational effect to
the particle. The third term $V_2=-E^2/r$ is the gravity of the
black hole to the particle. The Coulomb term $V_3=EQ^{*}/r$ acts
as a repulsive force effect to the particle for $Q^{*}>0$, but a
gravitational force effect for $Q^{*}<0$. The fifth term
$V_4=L^2/(2r^2\sin^2\theta)$ is an inertial centrifugal force
caused by the particle's angular momentum $L$. The sixth term
$V_5=Q^2E^2/(2r^2)$ is an electric field force, which acts as a
repulsive force effect to the particle. The magnetic field part
$V_1$ is a fundamental source for causing the nonintegrability and
chaoticity of the system (10). For $\beta=0$,  the system (10) is
integrable and nonchaotic. When $\beta$ is extremely small in the
case of $Q=0$ corresponding to $V_3=\emph{V}_5=0$, the black
hole's gravity $V_2$ is a dominant force and therefore chaos does
not possibly occur, either.  With $\beta$ increasing, the magnetic
field force increases. Only when $V_1$ appropriately matches with
$V_2$, may chaos occur. The extent of chaos can be strengthened
with an increase of the magnetic parameter from the global
phase-space structure. As to  $V_3$ and $V_5$ to the contributions
of particle's dynamics, $V_3\sim 1/r$ is a primary part, and
$V_5\sim 1/r^2$ is a secondary part for $r\gg2$ and $|Q|\leq 1$.
This can explain why the variation of $Q^{*}$ rather than the
variation of $Q$ is considerably sensitive to affect the regular
and chaotic dynamics of particles' orbits. For $Q^{*}>0$, the
Coulomb term, as a repulsive force, reduces the gravitational
effect from the magnetic field. On the contrary, the Coulomb term,
as a gravitational force, enhances the magnetic field
gravitational force effect. Thus, an increase of the magnitude of
negative Coulomb parameter $Q^{*}$ leads to strengthening the
extent of chaos, whereas  an increase of positive Coulomb
parameter $Q^{*}$ does not.

\section{Conclusions}

In this paper, we are devoted to designing explicit symplectic
integrators for a Hamiltonian system of charged test particles
moving around the Reissner-Nordstr\"{o}m black hole immersed into
an external magnetic field. In our construction, the Hamiltonian
is separated into five independently integrable parts with
analytical solutions as explicit functions of proper time. These
analytical solutions are used to yield second- and fourth-order
explicit symplectic integrators in symmetric combinations.

The proposed algorithms are shown to exhibit good long term
numerical performance in the Hamiltonian conservation, numerical
accuracy and computational efficiency. Such good numerical
performance does not mainly depend on the regular and chaotic
dynamical behavior of orbits but a step size. Thus, an optimal
step size is necessary.

The fourth-order explicit symplectic integrator with an optimal
step size is applied to well explore the dynamics of charged
particles around the Reissner-Nordstr\"{o}m black hole with an
external magnetic field. We focus on the influences of the black
hole's charge, the Coulomb part of the electromagnetic potential
and the magnetic parameter on the dynamical behavior. The magnetic
parameter plays an important role in causing the nonintegrability
and chaoticity of the system. Under some circumstances, the extent
of chaos is strengthened from the global phase-space structure as
the magnetic parameter increases. No the variation of the black
hole's charge but the variation of the Coulomb part is
considerably sensitive to affect the regular and chaotic dynamics
of particles' orbits. A positive Coulomb part is easier to induce
chaos than a negative one.

\section*{Acknowledgments}

The authors are very grateful to a referee for useful suggestions.
This research has been supported by the National Natural Science
Foundation of China [Grant Nos. 11533004, 11973020 (C0035736),
11803020, 41807437, U2031145]
and the Natural Science Foundation of Guangxi (Grant Nos.
2018GXNSFGA281007 and 2019JJD110006).

\begin{figure*}[ptb]
\center{
\includegraphics[scale=0.25]{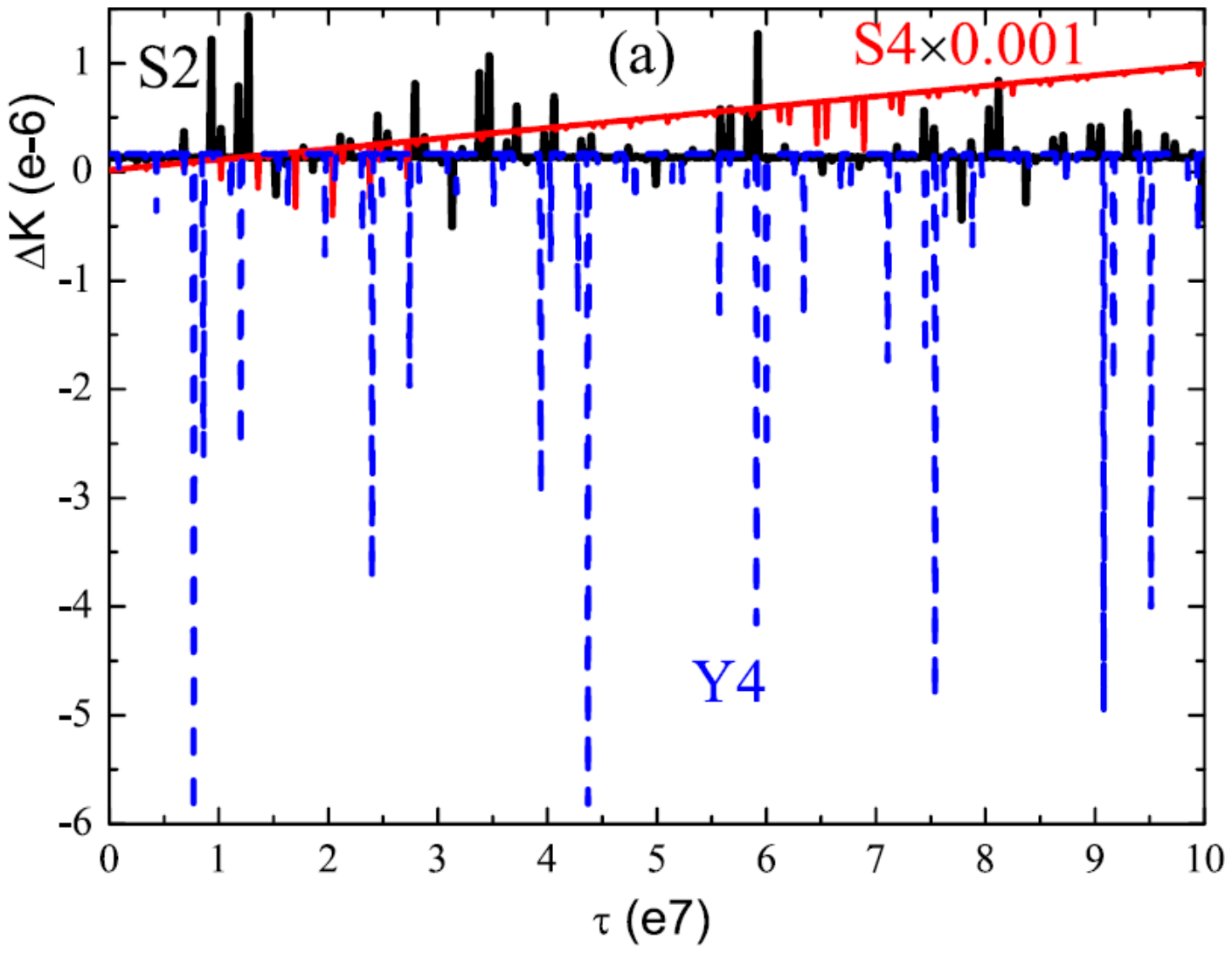}
\includegraphics[scale=0.25]{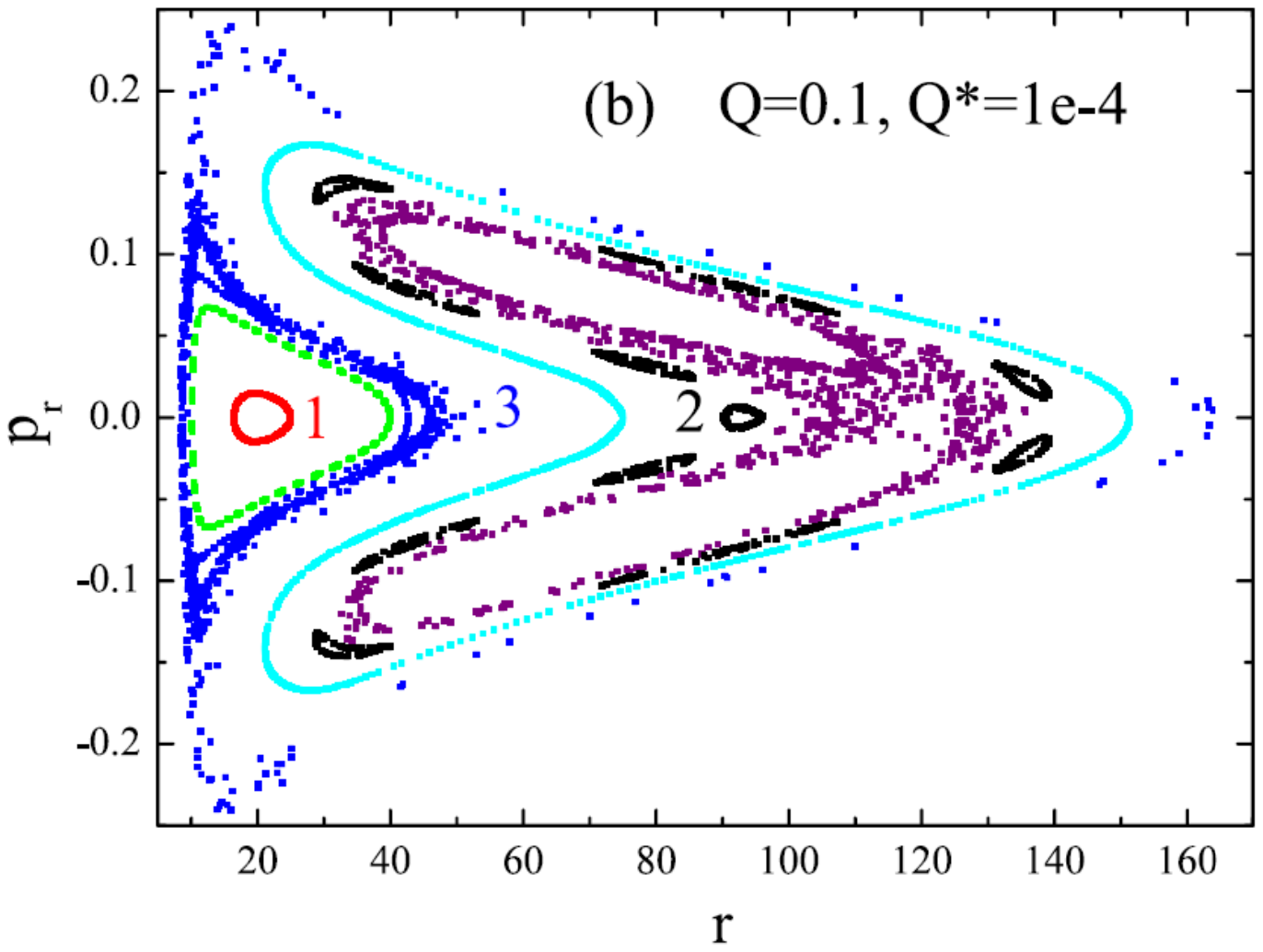}
\caption{(a) Hamiltonian errors $\Delta K=-1/2-K$ in Equation (13)
for the proposed algorithms solving the system (10). The
parameters are $E=0.995$, $L=4.6$, $\beta=6.4\times10^{-4}$,
$Q=0.1$ and $Q^{*}=1\times10^{-4}$. A test orbit has initial
conditions $r=25$, $p_r=0$ and $\theta=\pi/2$. The starting value
of  $p_{\theta}>0$ is determined by Equation (13). The new
second-order and fourth-order explicit symplectic integrators S2
and S4 take proper time step $h=1$. The  realistic errors for S4
are 1000 times smaller than the plotted errors. The errors for S2
remain bounded and stable in an order of $\mathcal{O}(10^{-6})$,
whereas do not for S4 due to roundoff errors. However, such a
secular drift in Hamiltonian errors is missing when large proper
time step $h=10$ is used in the fourth-order method Y4. These
facts show that the new algorithms with appropriate time steps can
share good properties of a standard symplectic integrator in
long-term stabilized error behavior. (b) Poincar\'{e} sections on
the plane $\theta=\pi/2$ and $p_{\theta}>0$, given by algorithm S2
with proper time step $h=1$. The test orbit in panel (a) is
regular Orbit 1 colored red in panel (b).}
 \label{Fig1}}
\end{figure*}

\begin{figure*}[ptb]
\center{
\includegraphics[scale=0.18]{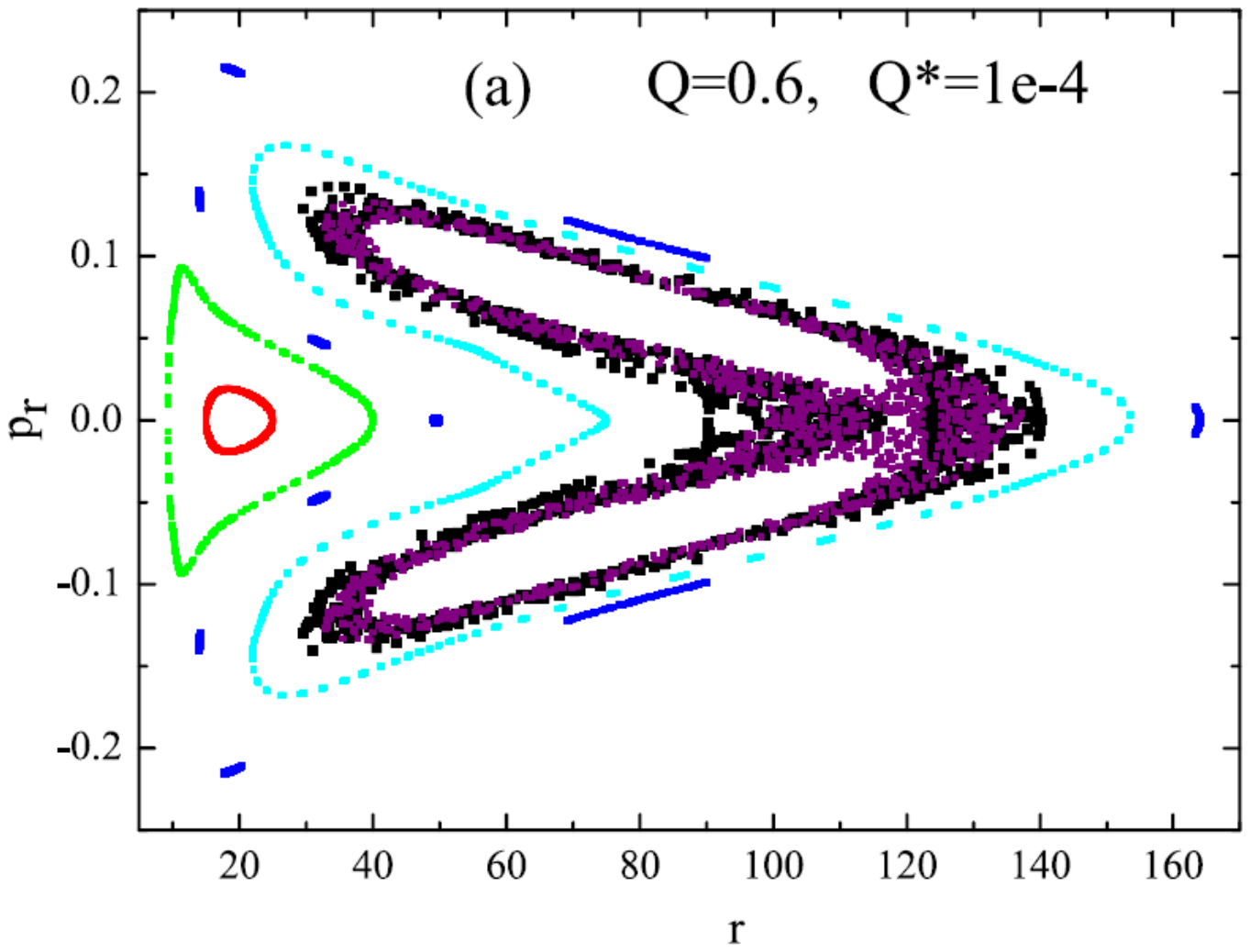}
\includegraphics[scale=0.18]{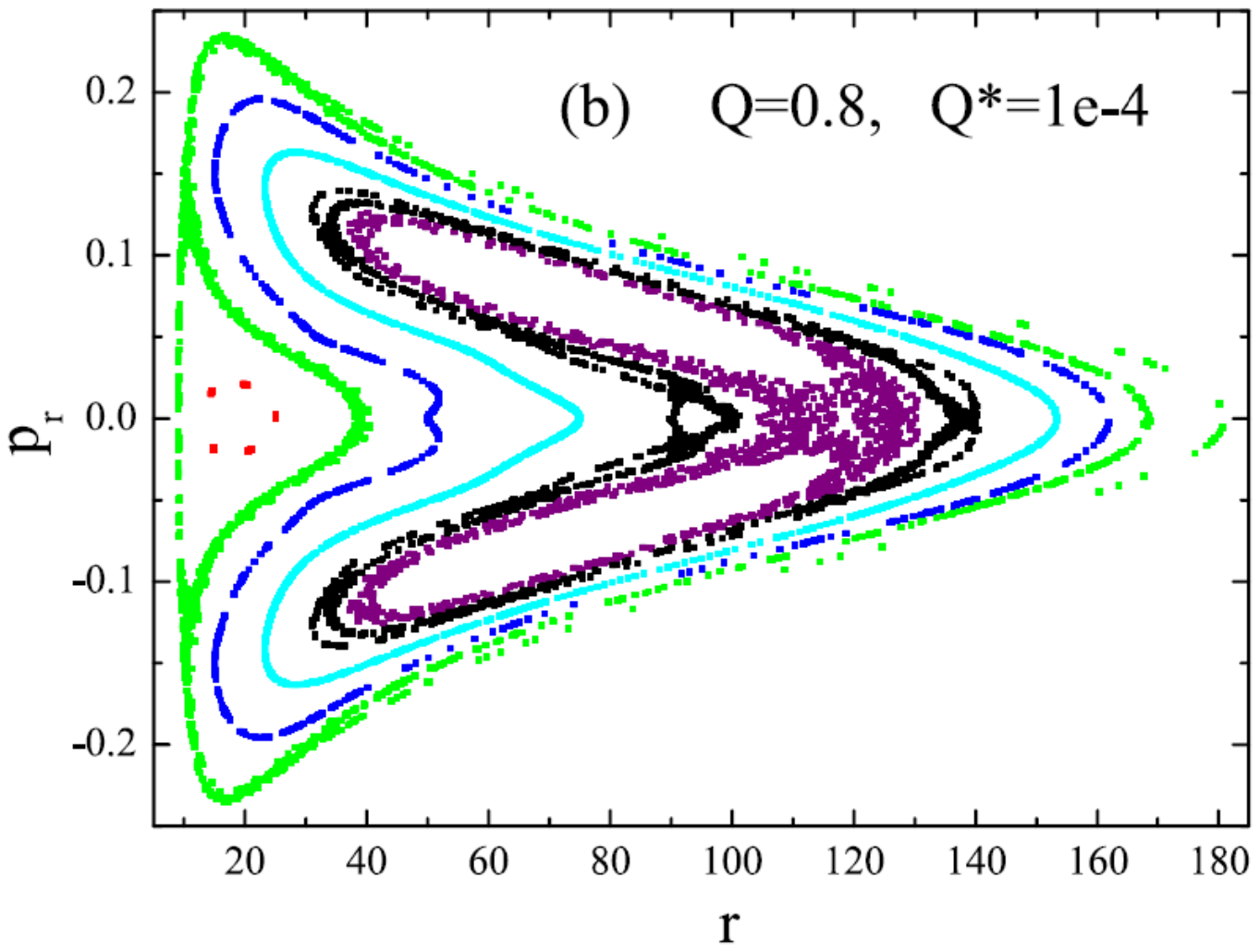}
\includegraphics[scale=0.18]{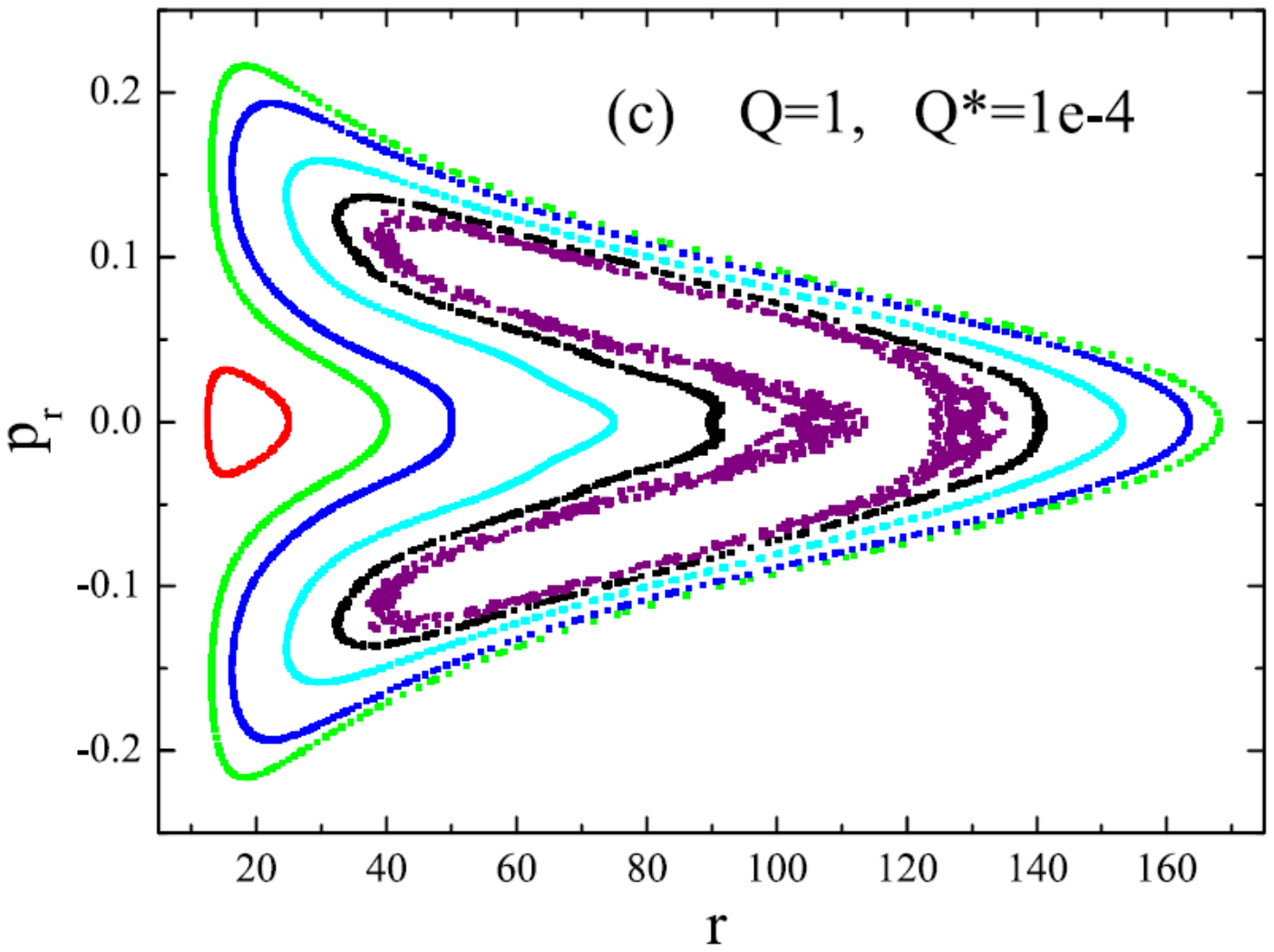}
\includegraphics[scale=0.18]{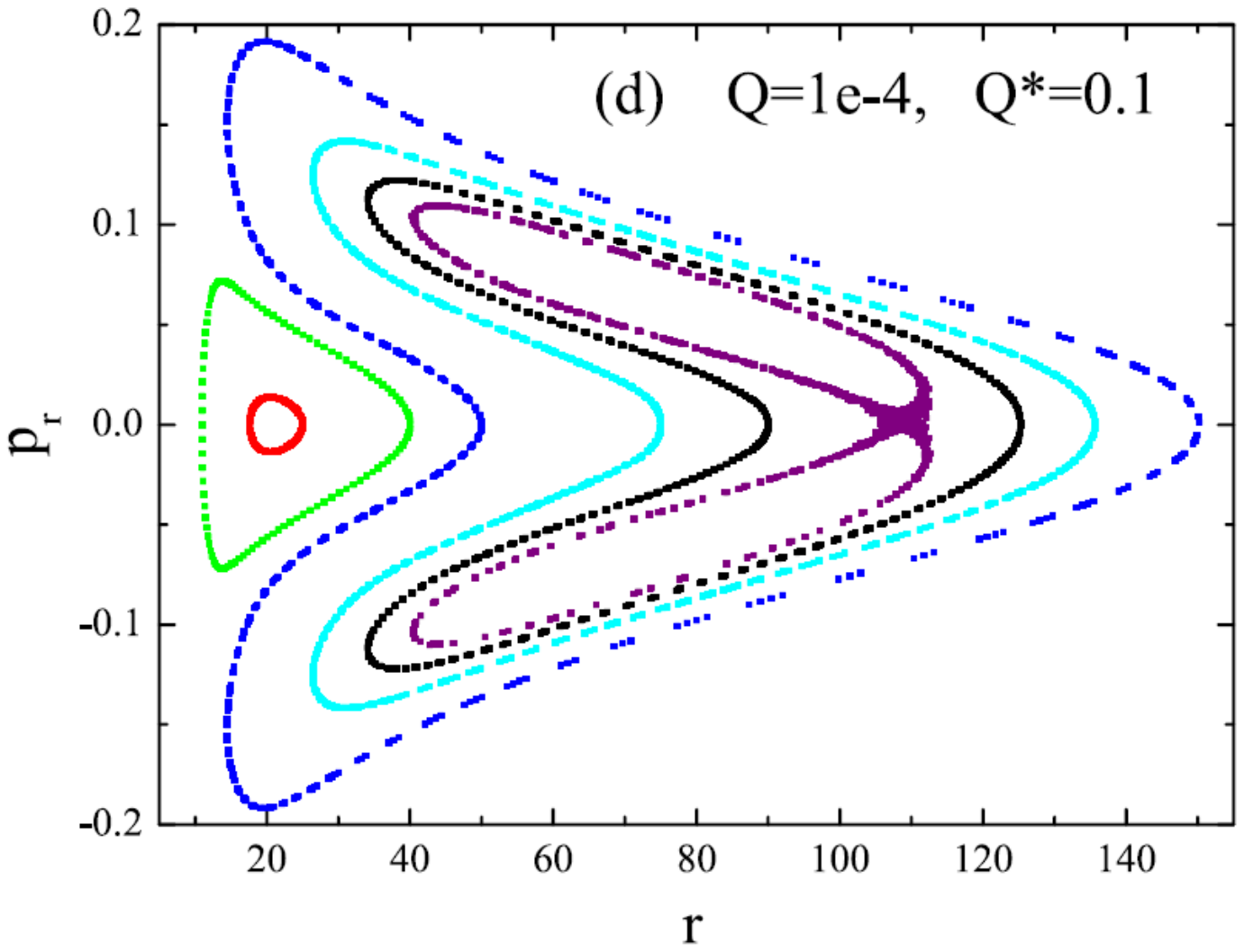}
\includegraphics[scale=0.18]{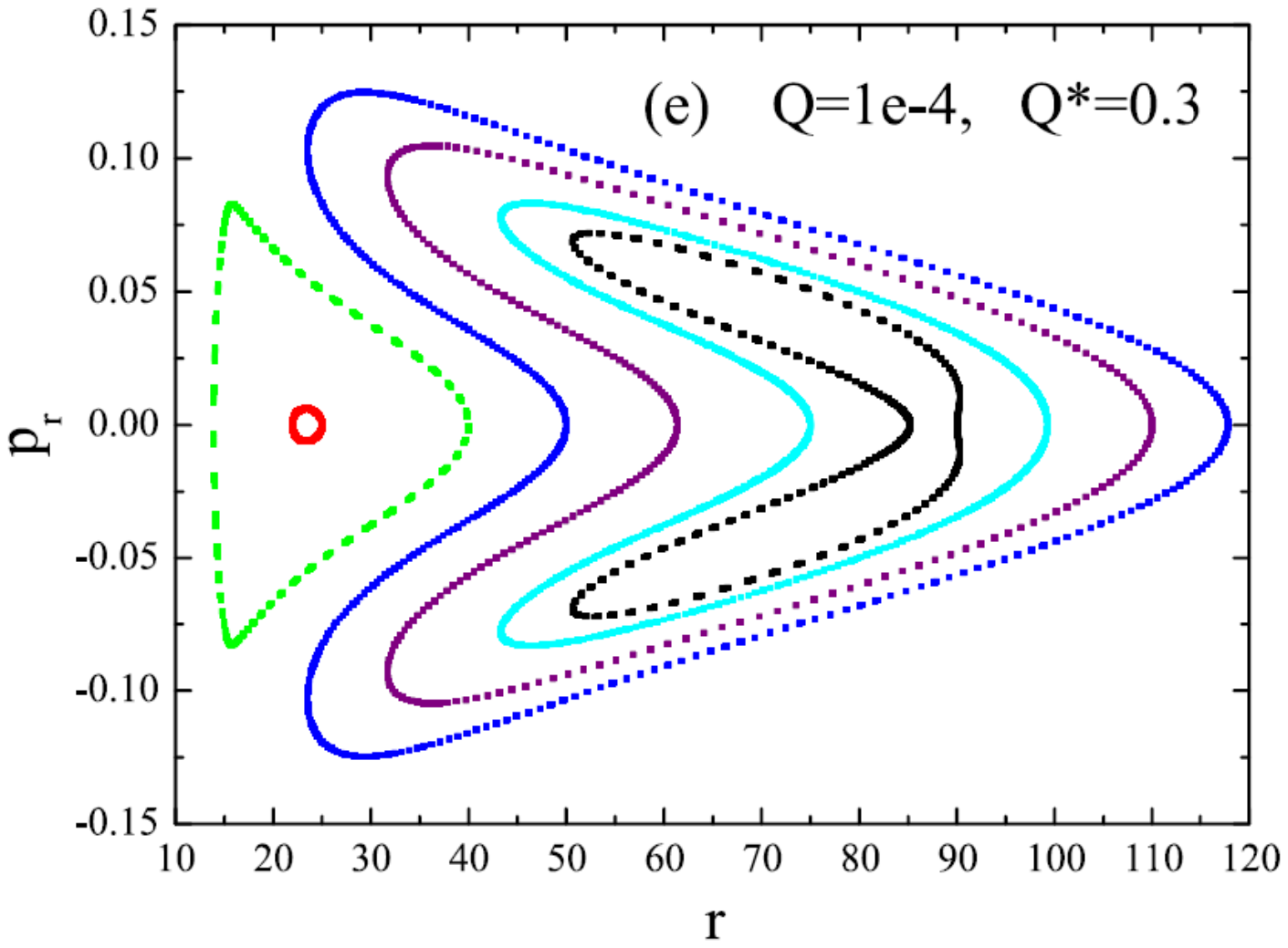}
\includegraphics[scale=0.18]{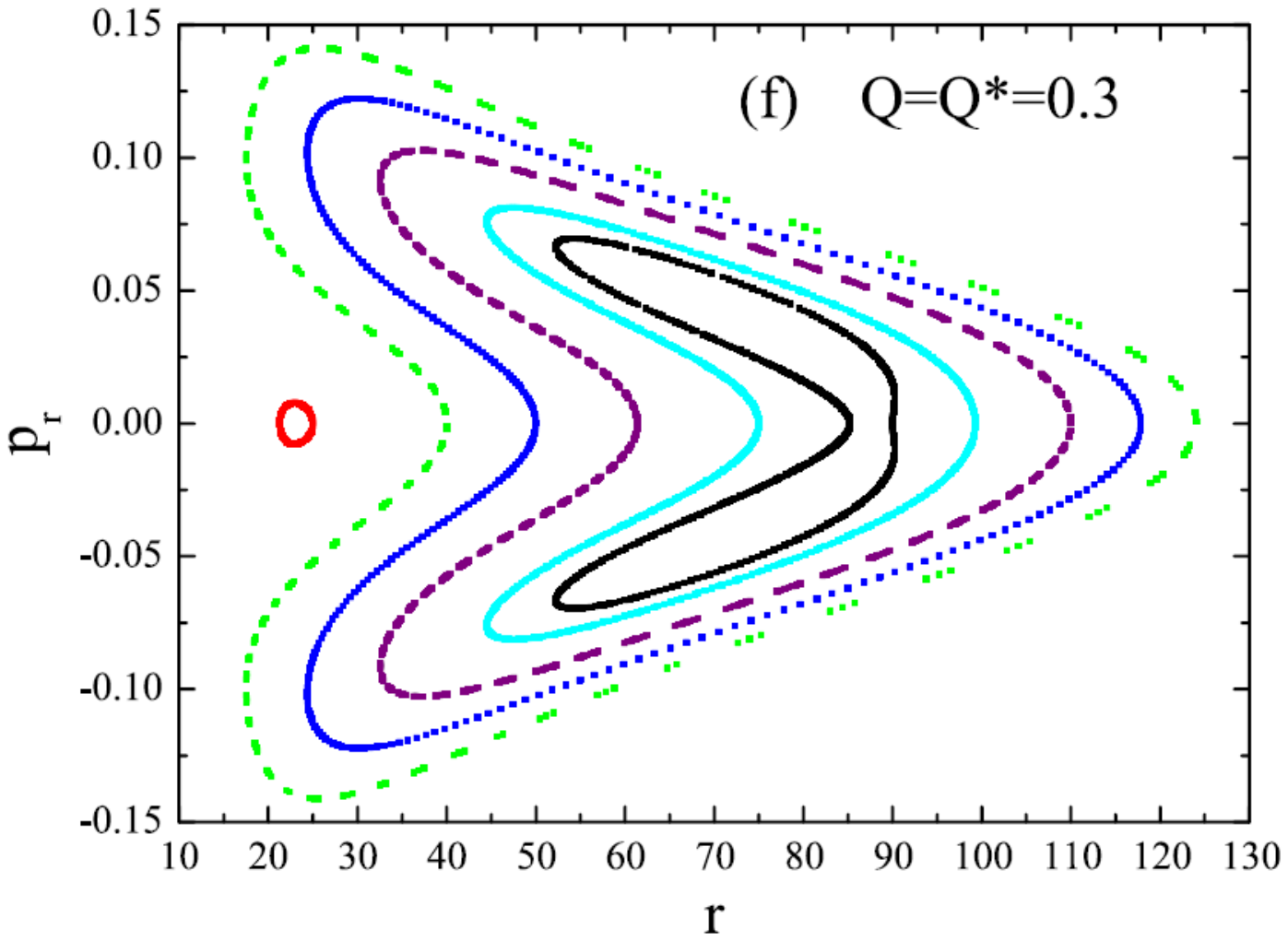}
\includegraphics[scale=0.18]{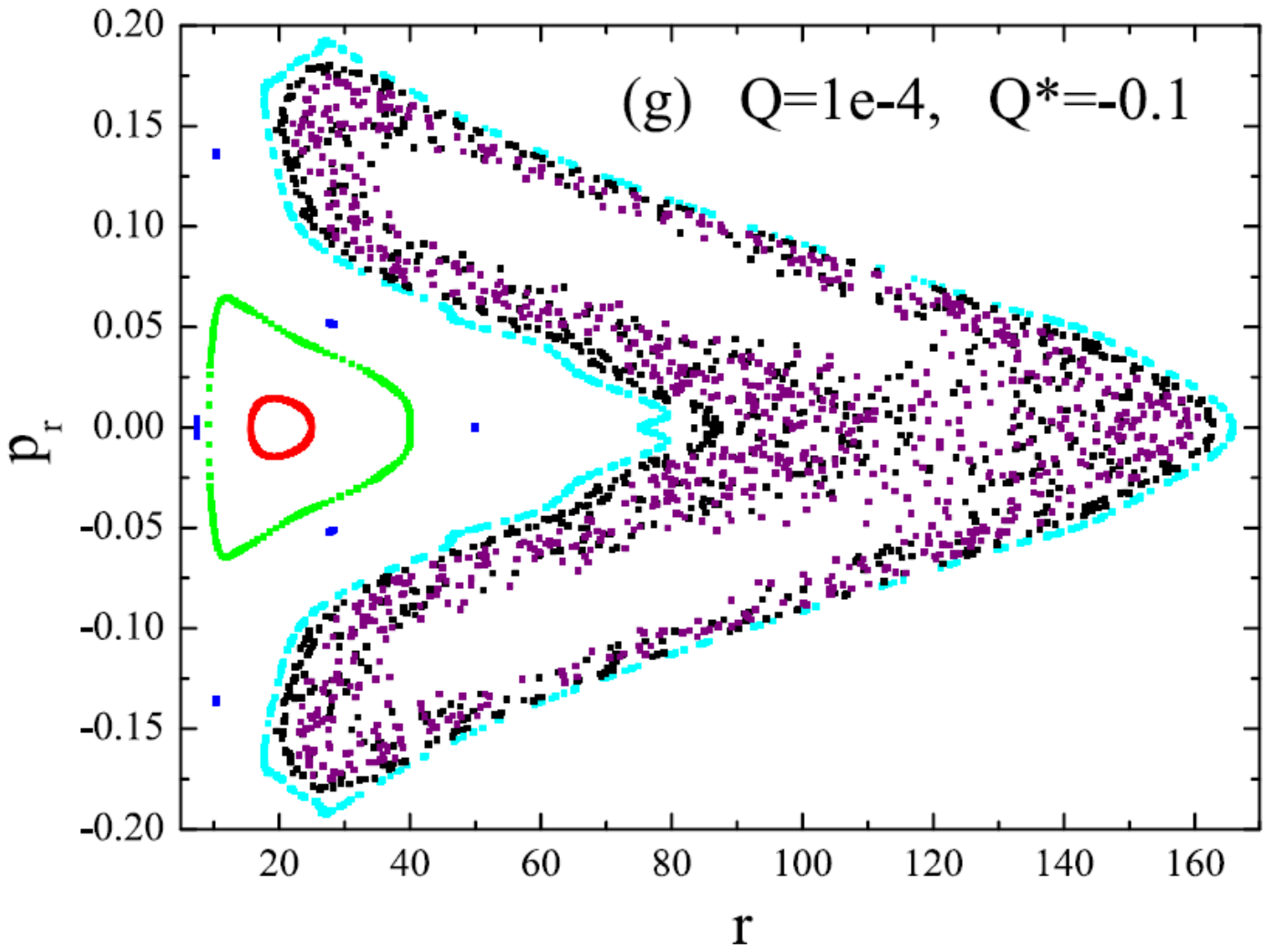}
\includegraphics[scale=0.18]{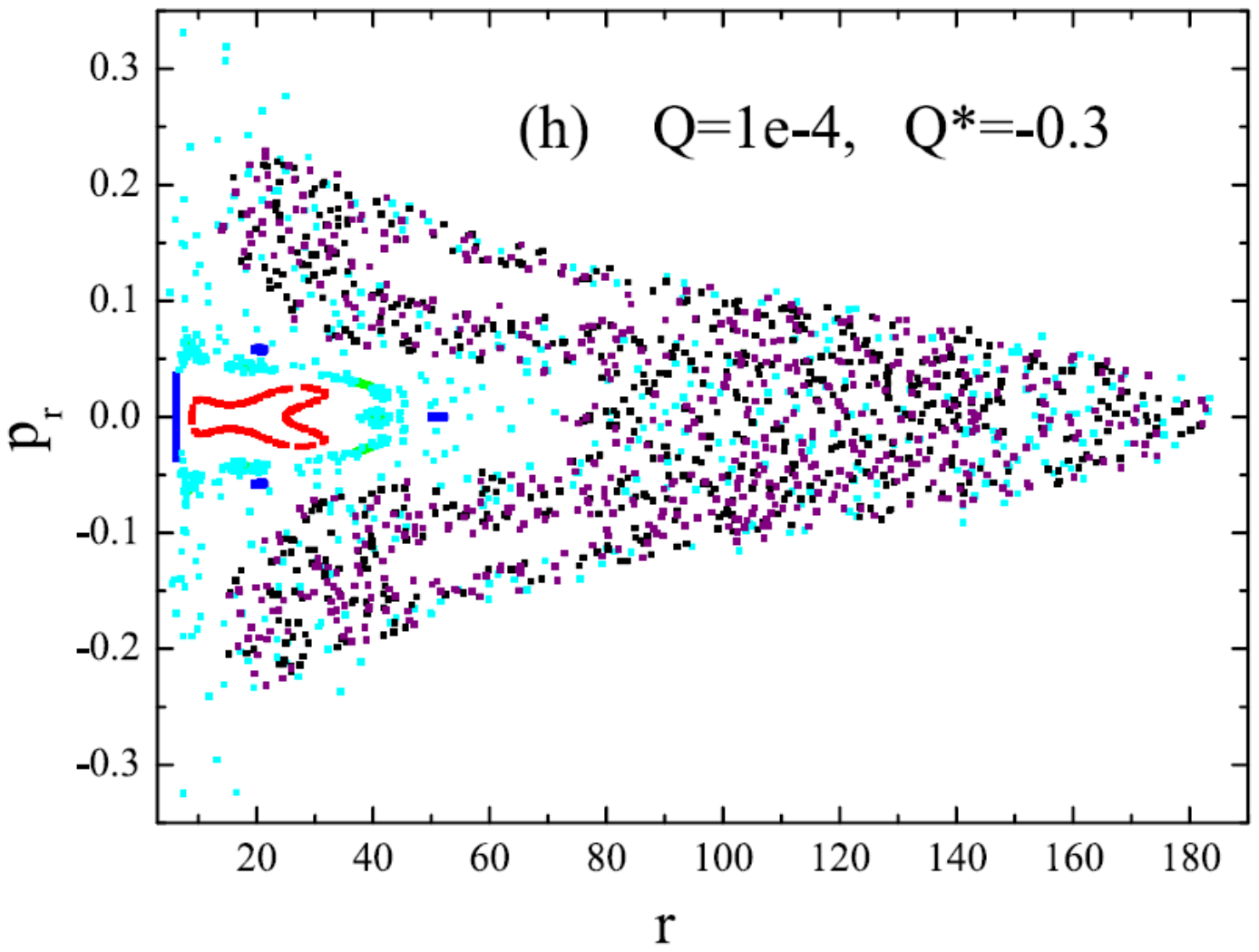}
\includegraphics[scale=0.18]{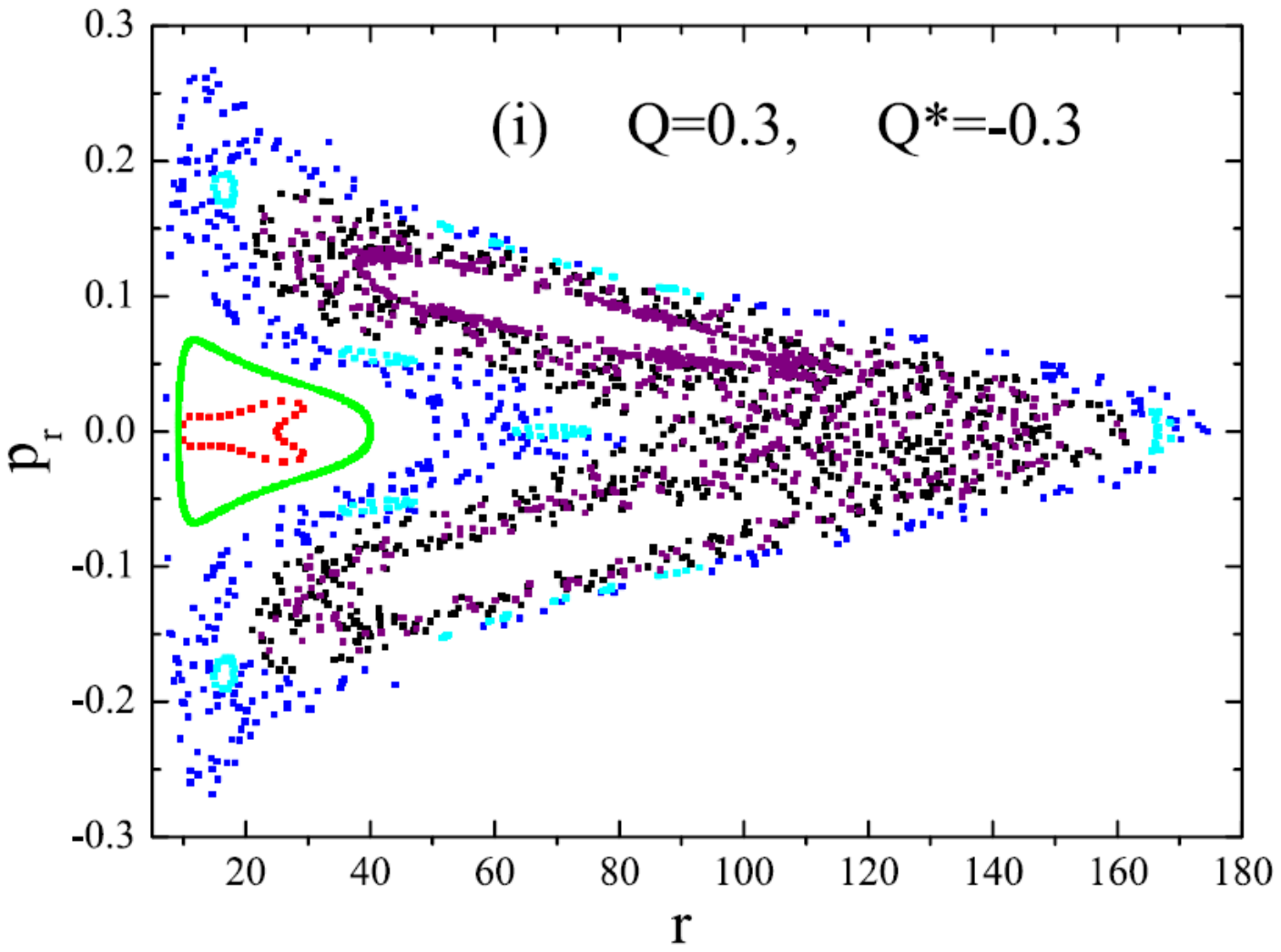}
\caption{Same as Figure 1(b), but different values of parameters
$Q$ and $Q^{*}$ are given.}
 \label{Fig2}}
\end{figure*}

\begin{figure*}[ptb]
\center{
\includegraphics[scale=0.18]{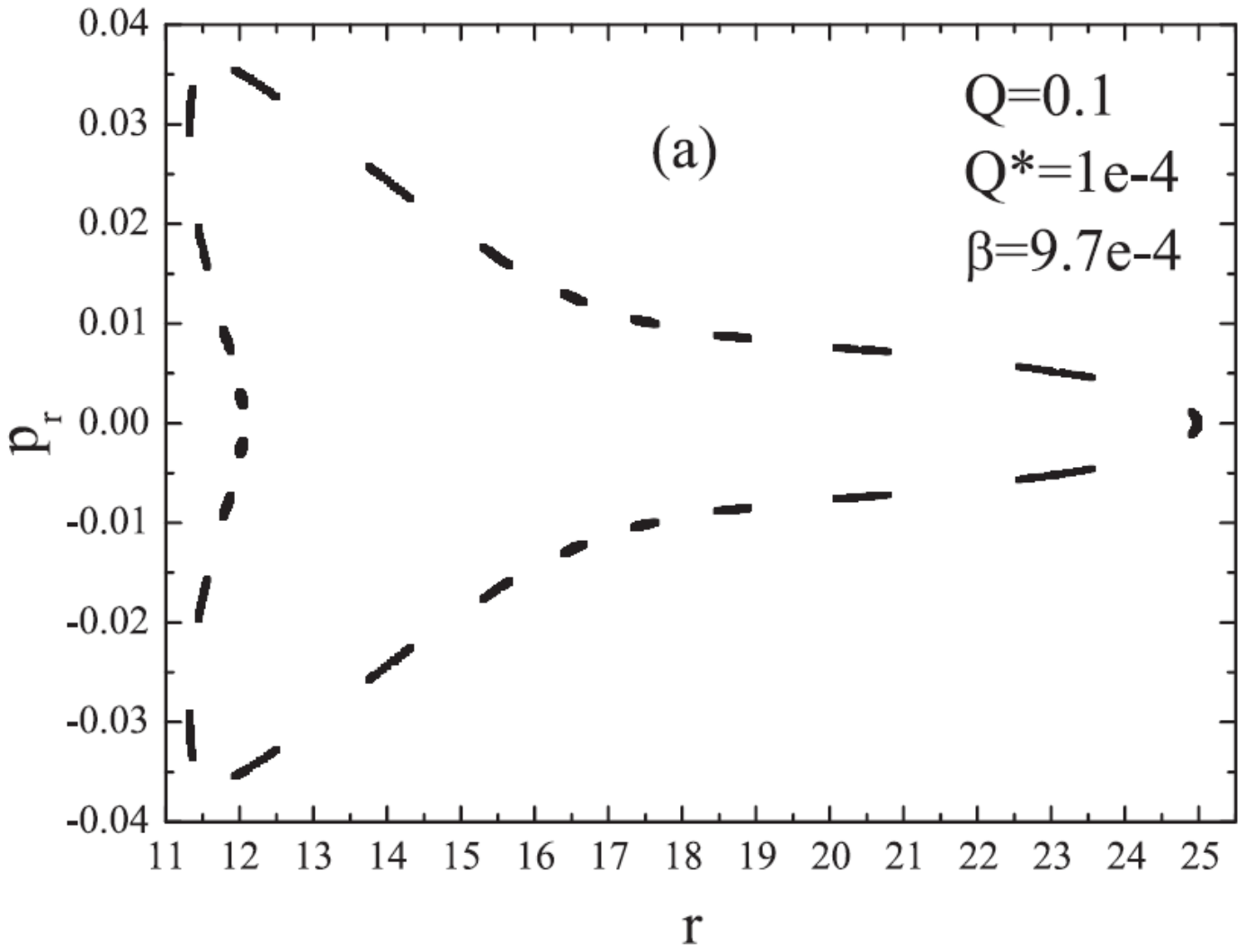}
\includegraphics[scale=0.18]{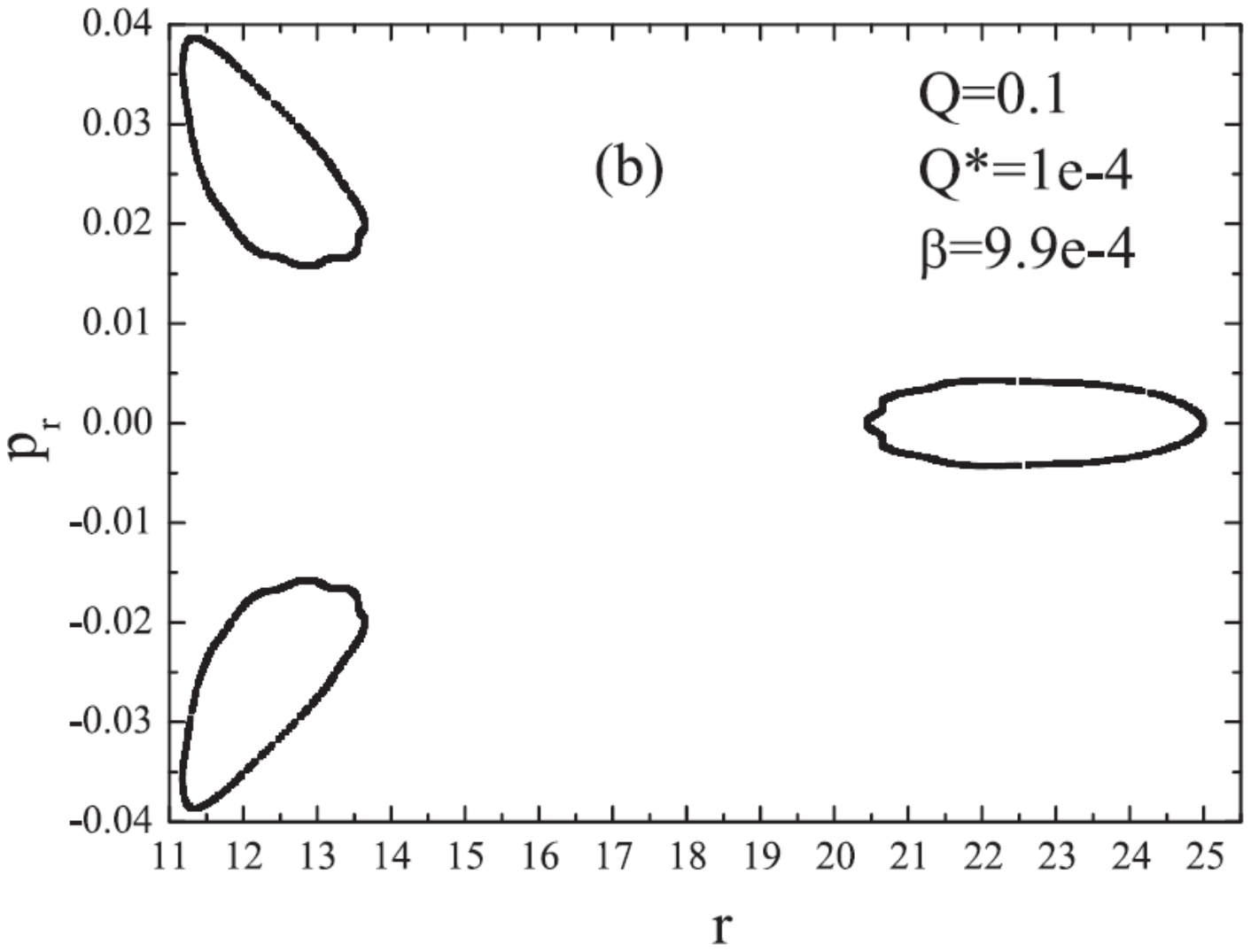}
\includegraphics[scale=0.18]{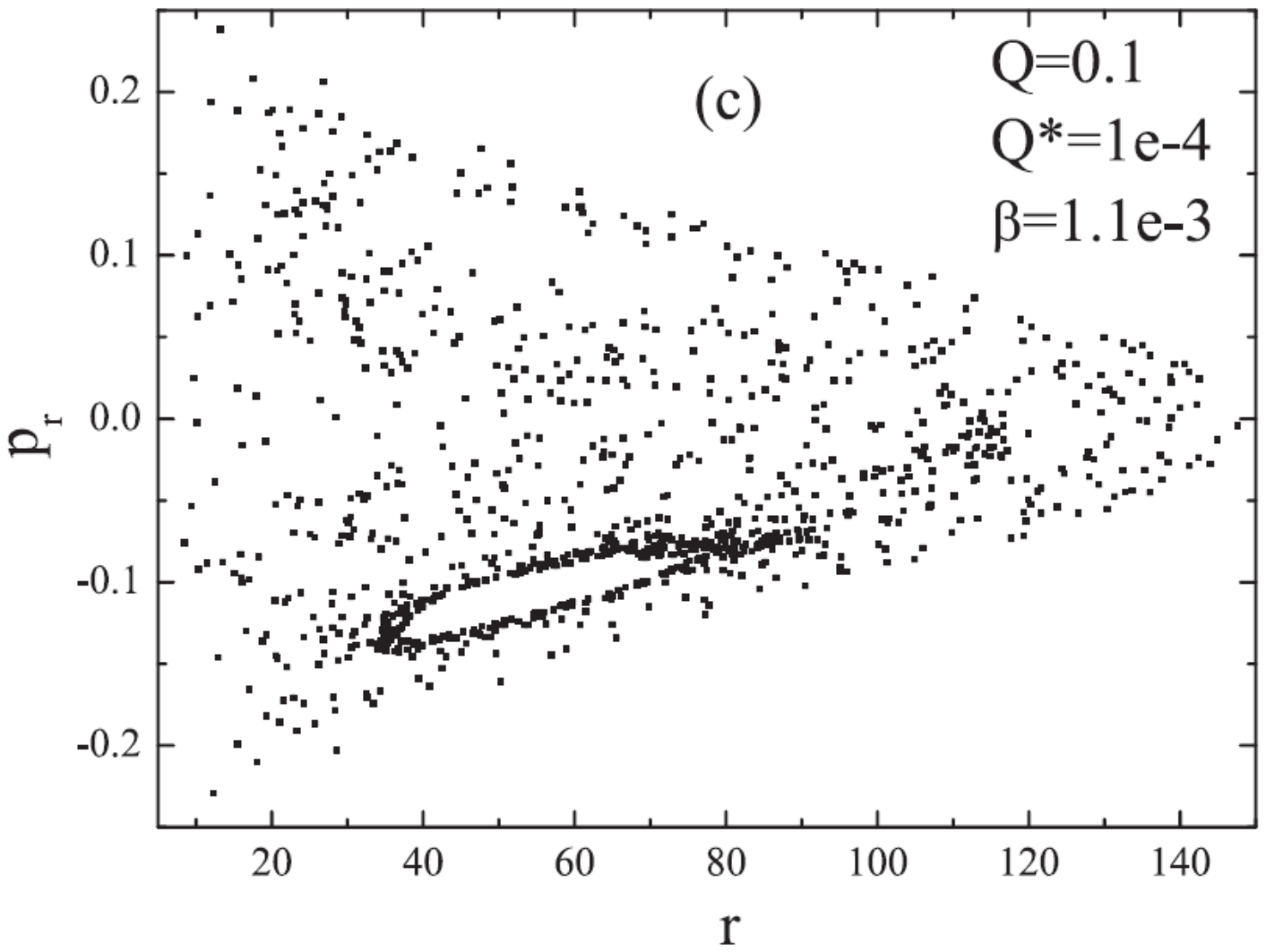}
\includegraphics[scale=0.18]{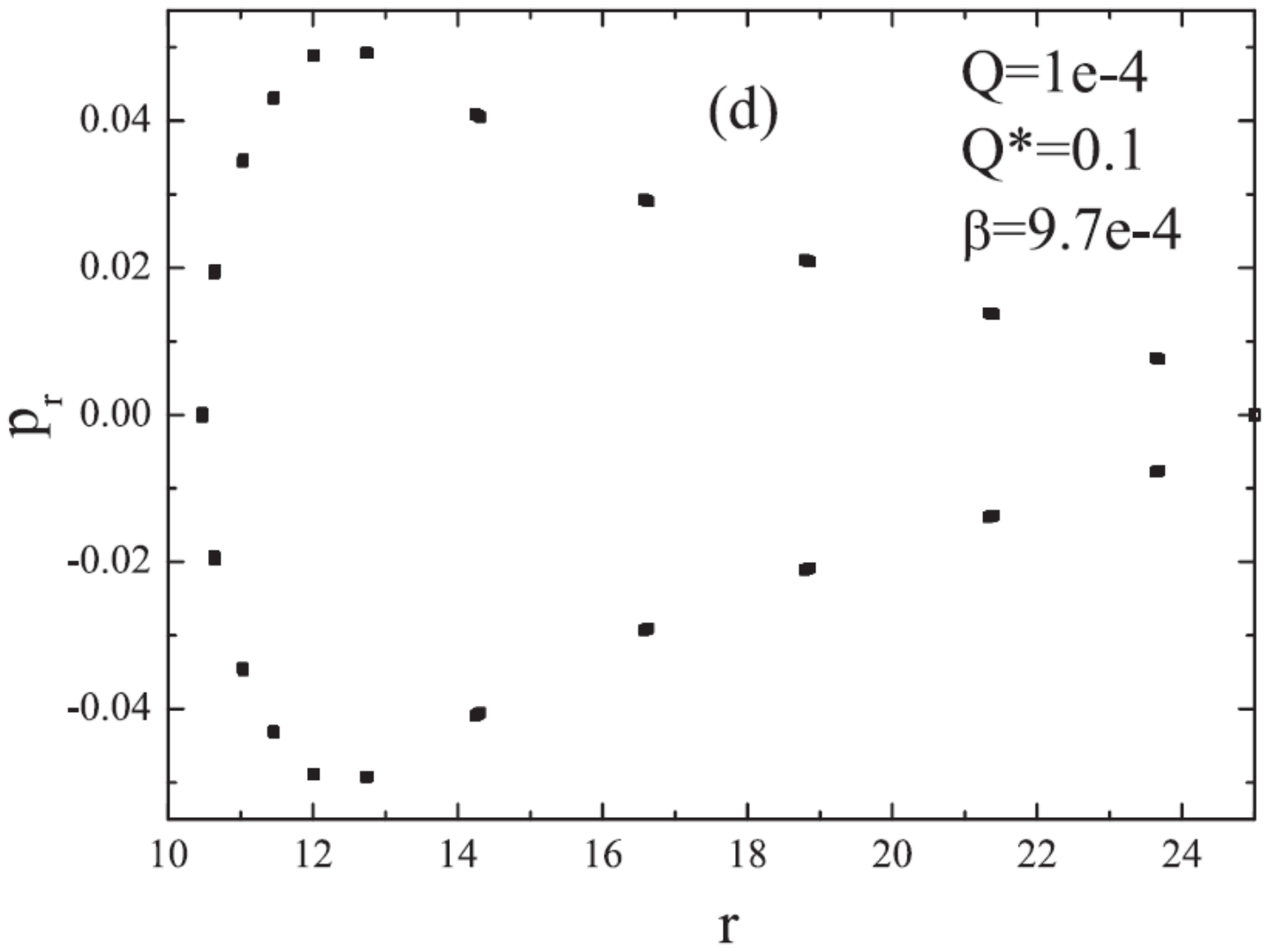}
\includegraphics[scale=0.18]{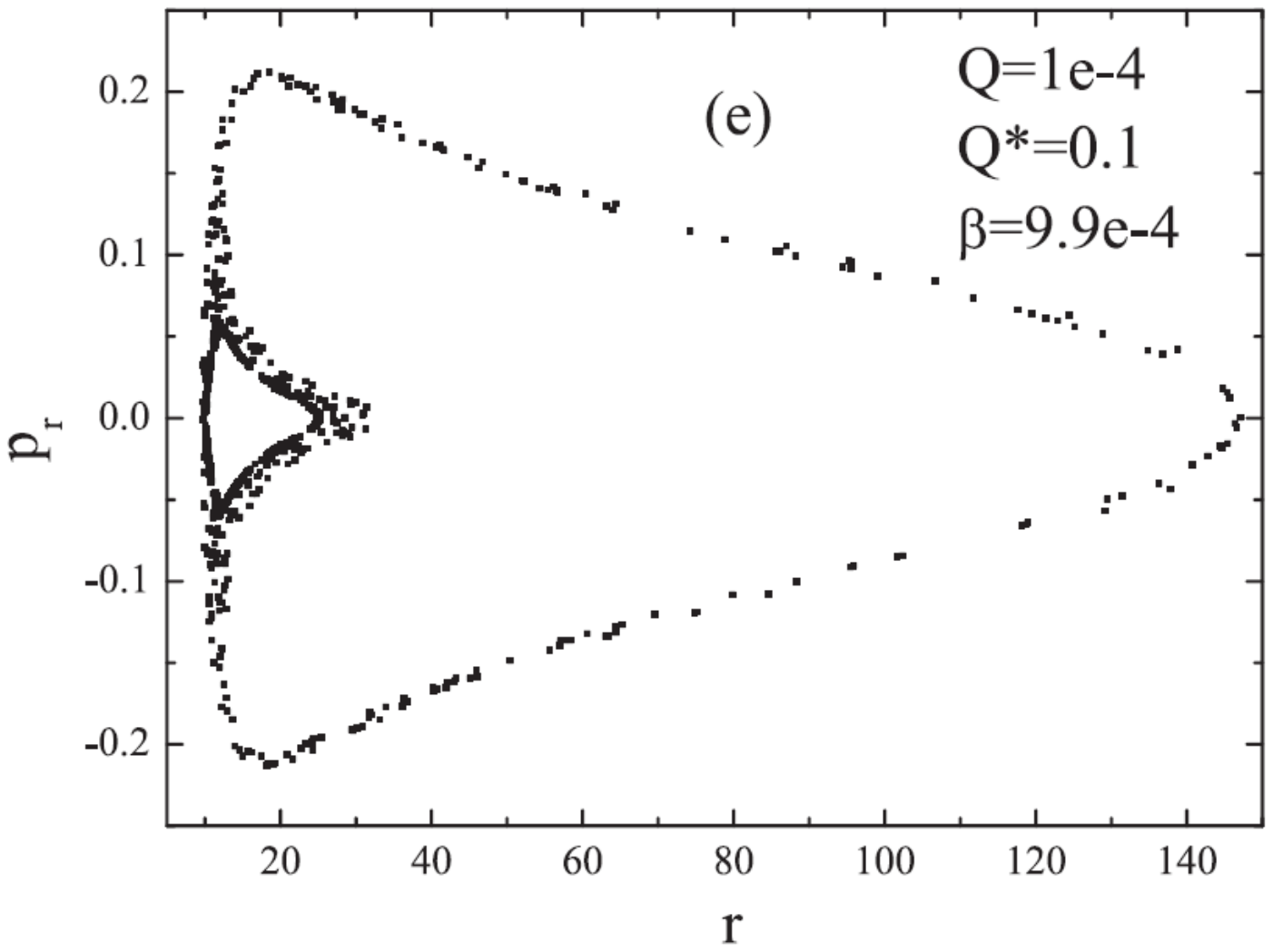}
\includegraphics[scale=0.18]{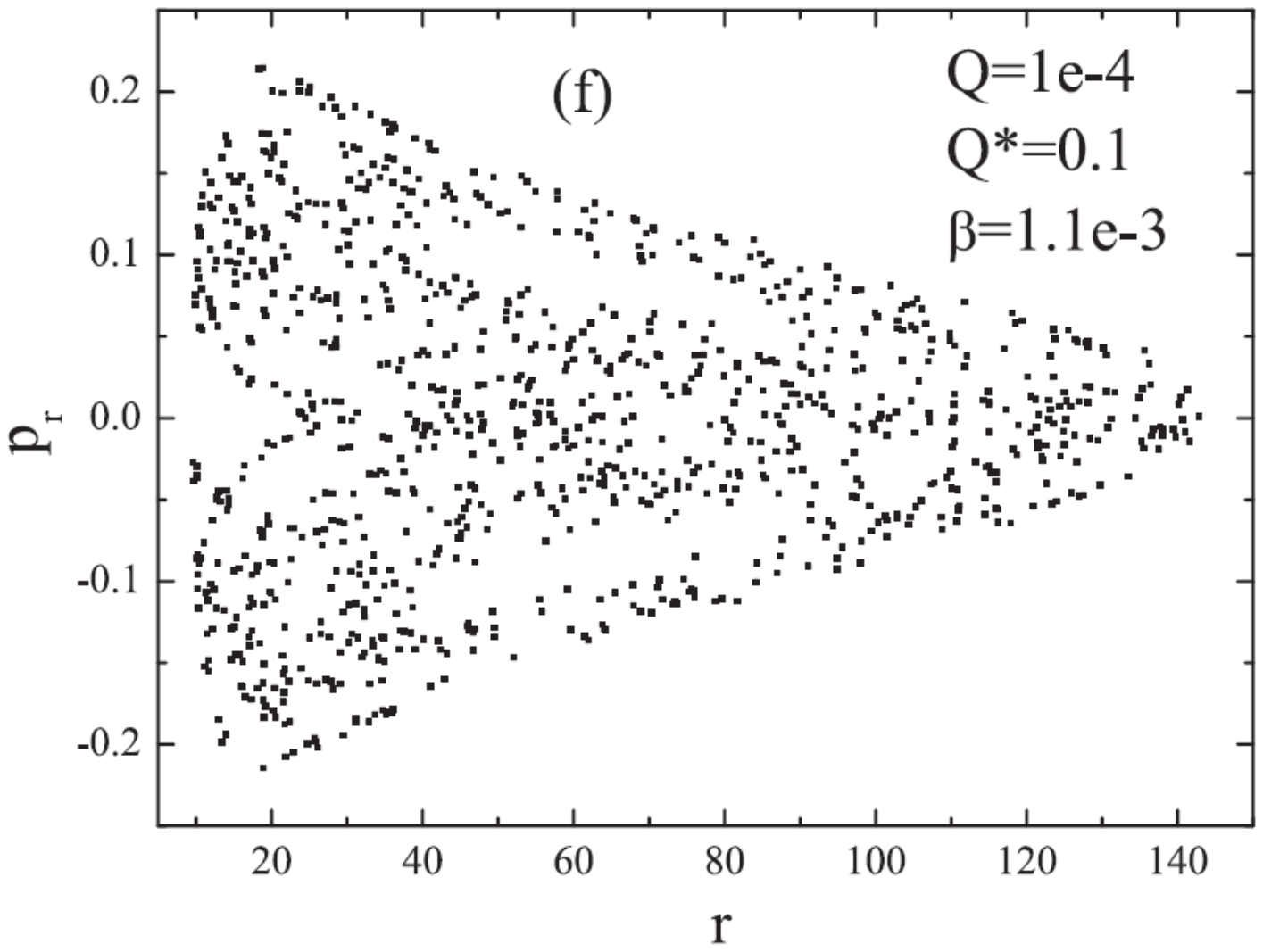}
\includegraphics[scale=0.18]{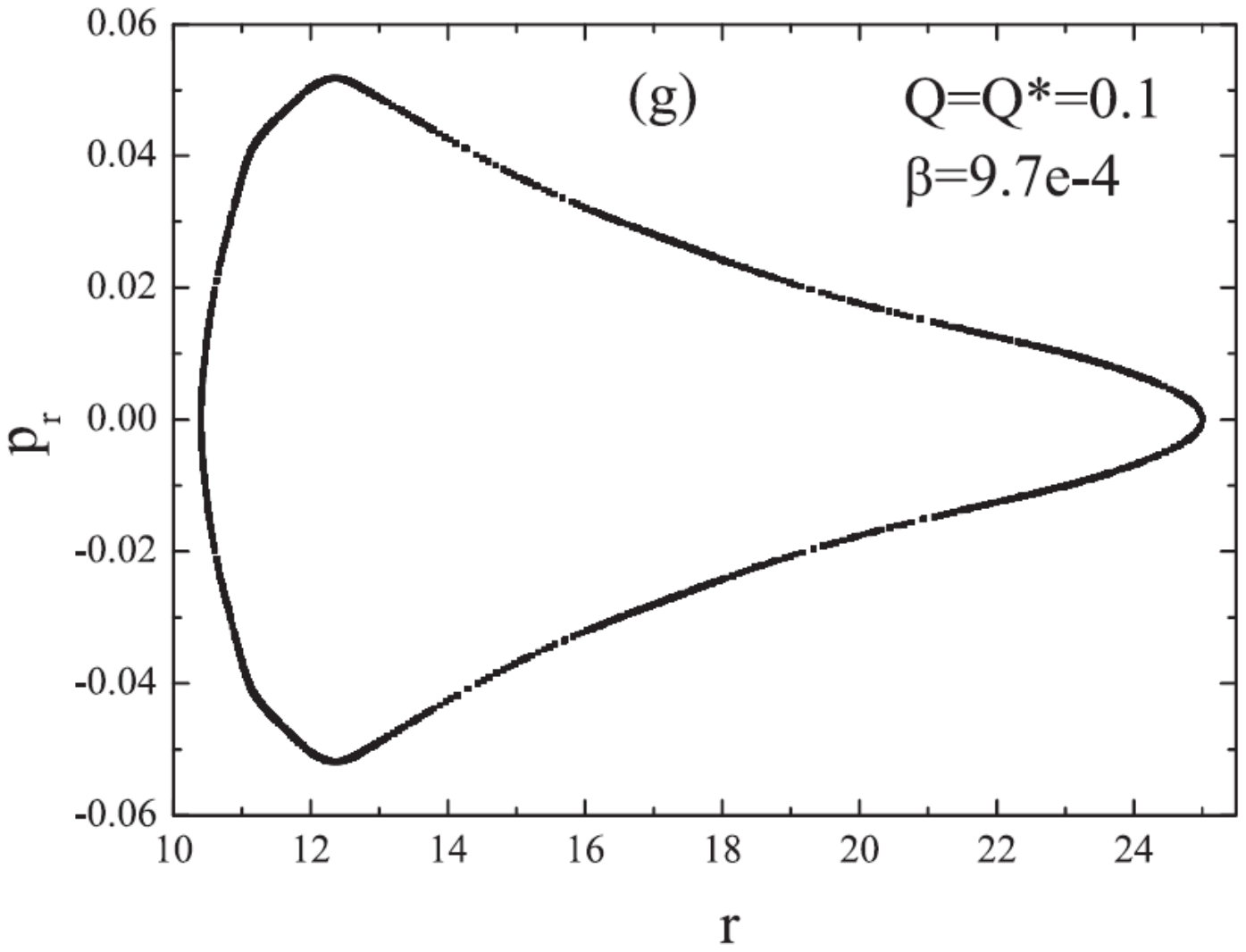}
\includegraphics[scale=0.18]{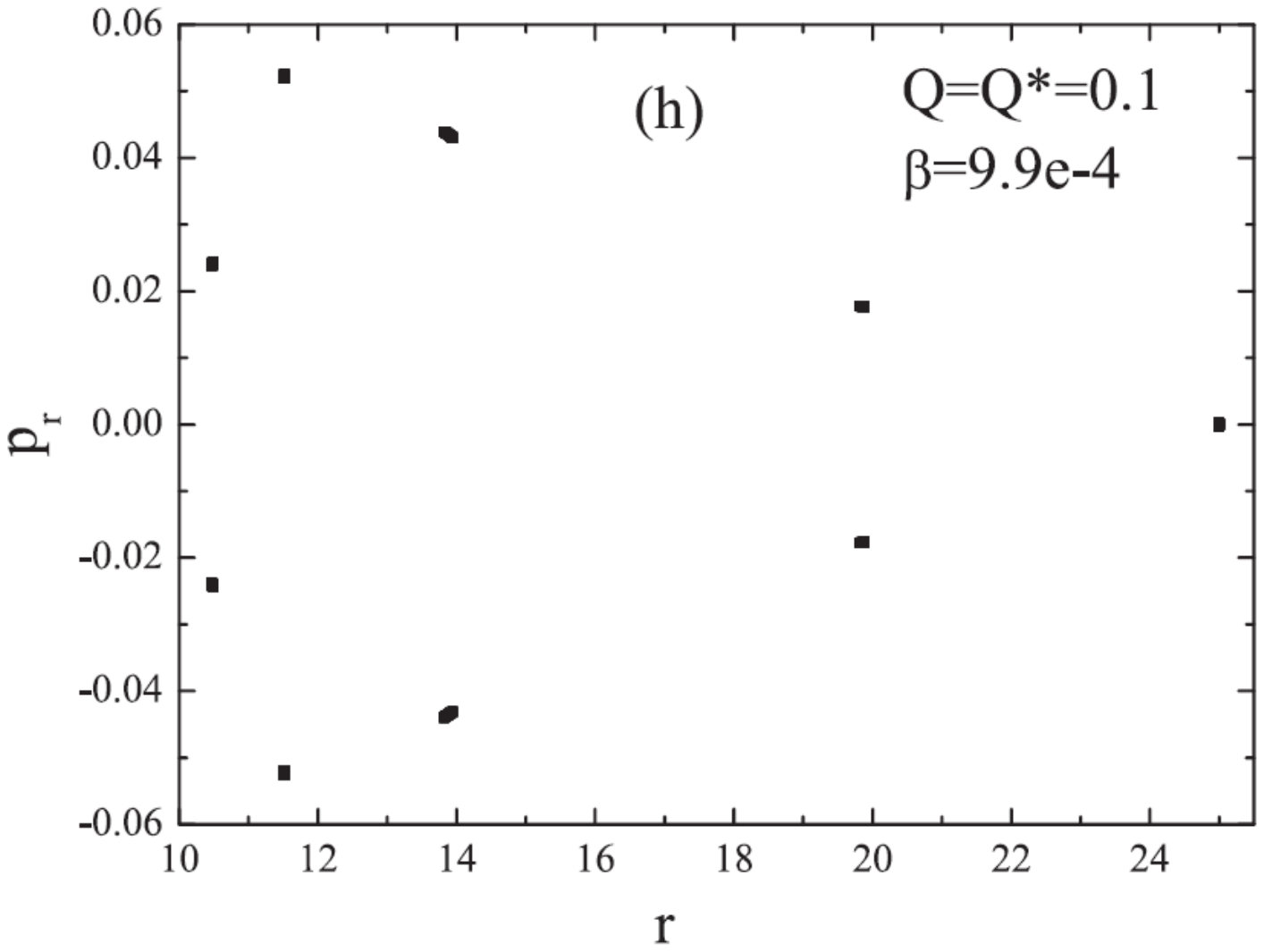}
\includegraphics[scale=0.18]{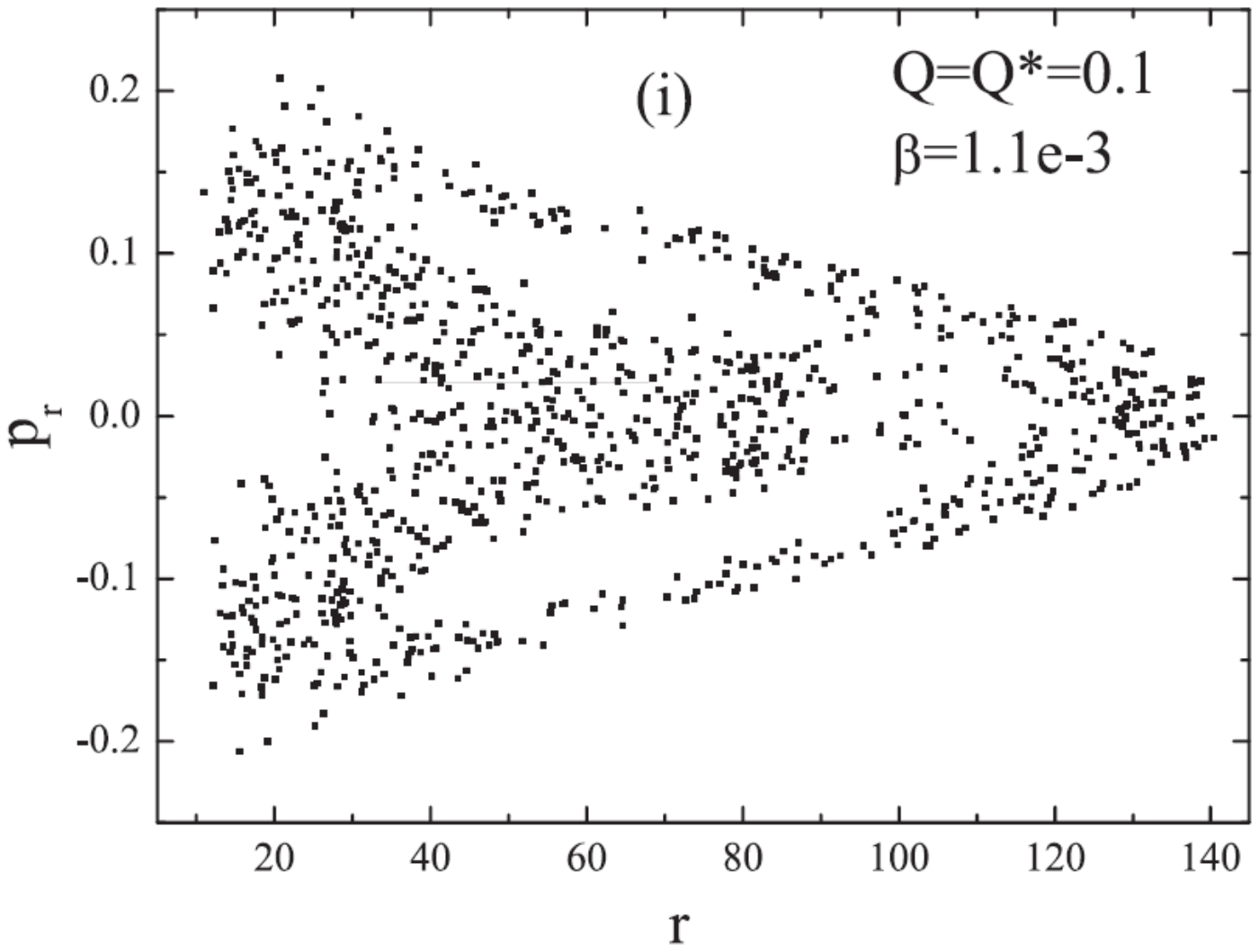}
\caption{Same as Orbit 1 colored red in Figure 1(b), but different
combinations of parameters $Q$, $Q^{*}$ and $\beta$ are given.}
 \label{Fig3}}
\end{figure*}

\end{document}